\documentclass[11pt]{article}
\usepackage{CJK}
\usepackage{graphicx}
\usepackage{fancyhdr}
\usepackage{amsmath,amsthm,amssymb}
\usepackage{bm}
\usepackage{mathrsfs}
\usepackage{multirow}
\usepackage{indentfirst}
\usepackage{url}
\usepackage{stmaryrd}
\usepackage{subcaption}
\usepackage{color}
\usepackage{epstopdf}
\usepackage{natbib}
\usepackage[caption=false]{subfig}
\usepackage[utf8]{inputenc}
\usepackage{epsfig}

\newtheorem{theorem}{Theorem}[section]
\newtheorem{remark}{Remark}
\newtheorem{lemma}{Lemma}[section]

\numberwithin{equation}{section}
\makeatletter
\newcommand{\Rmnum}[1]{\expandafter\@slowromancap\romannumeral #1@}
\makeatother
\newtheorem{corollary}{Corollary}[section]
\allowdisplaybreaks

\begin{document}

\title{\bf A Two-Step Projection-Based Goodness-of-Fit Test for Ultra-High Dimensional Sparse Regressions 
\footnote{Corresponding author. 
}
}
\author{Falong Tan$^{1}$, Jie Liu$^1$, Heng Peng$^{2*}$, and Lixing Zhu$^{3*}$ \\~\\
{\small {\small {\it $^1$ Department of Statistics and Data Science, Hunan University, Changsha, China}}}\\
{\small {\small {\it $^2$ Department of Mathematics, Hong Kong Baptist University, Hong Kong, China}}}\\
{\small {\small {\it $^3$ Department of Statistics, Beijing Normal University at Zhuhai, Zhuhai, China}}}
}
\date{}
\maketitle

\begin{abstract}
This paper proposes a novel two-step strategy for testing the goodness-of-fit of parametric regression models in ultra-high dimensional sparse settings, where the predictor dimension far exceeds the sample size. This regime usually renders existing goodness-of-fit tests for regressions infeasible, primarily due to the curse of dimensionality or their reliance on the asymptotic linearity and normality of parameter estimators---properties that may no longer hold under ultra-high dimensional settings. To address these limitations, our strategy first constructs multiple test statistics based on projected predictors from distinct projections and establishes their asymptotic properties under both the null and alternative hypotheses. This projection-based approach significantly mitigates the dimensionality problem, enabling our tests to detect local alternatives converging to the null at the rate as if the predictor were univariate. An important finding is that the resulting test statistics based on linearly independent projections are asymptotically independent under the null hypothesis. Based on this, our second step employs powerful $p$-value combination procedures, such as the minimum $p$-value and the Fisher combination of $p$-value, to form our final tests and enhance power. Theoretically, our tests only require the standard convergence rate of parameter estimators to derive their limiting distributions, thereby circumventing the need for asymptotic linearity or normality of parameter estimators. Simulations and real-data applications confirm that our approach provides robust and powerful goodness-of-fit testing in ultra-high dimensional settings. \\

\end{abstract}

\newpage


\setcounter{equation}{0}
\section{Introduction}
Over the last three decades, substantial research has focused on developing estimation methodologies for high-dimensional regression models, where the predictor dimension $p$ can far exceed the sample size $n$ \citep{buhlmann2011}. To avoid unreliable conclusions, any statistical analysis employing such models must also include a model checking procedure. However, this critical area has not been systematically explored in ultra-high dimensional settings \citep{shao2018}. In this paper, we aim to develop goodness-of-fit tests for regression models when the predictor dimension $p$ may substantially exceed the sample size $n$. Consider the following model:
\begin{equation}\label{nonpara-model}
    Y=m(X) + \varepsilon,
\end{equation}
where $Y$ is the response variable, $X \in \mathbb{R}^p$ is the predictor vector, and $\varepsilon=Y-E(Y|X)$ is the error term. Our goal is to test whether $(Y, X)$ adheres to a specified parametric regression model $ Y=m(X, \beta_0) + \varepsilon(\beta_0)$ with $E[\varepsilon(\beta_0)|X] = 0$ for some unknown parameter $\beta_0$, in ultra-high dimensional settings---such as ultra-high dimensional linear models or generalized linear models.

In the area of model checking, how to handle the dimensionality of predictors is often a significant challenge. In low-dimensional settings ($p$ fixed), a primary class of tests for model checking is the empirical process-based tests. Examples include the works of \cite{stute1997}, \cite{Stute1998a}, \cite{Stute1998b}, \cite{zhu2003}, \cite{khmaladze2004}, \cite{escanciano2006b}, \cite{stute2008}, \cite{cuesta2019}. This class of test, without further transformation of their empirical processes---such as martingale transformation \citep{Stute1998b} or Neyman-orthogonal projection \citep{escanciano2024}---typically requires asymptotic linearity or normality of parameter estimators to derive the limiting distributions. However, these properties may not hold in ultra-high dimensional scenarios, making a direct extension of corresponding empirical process-based tests to ultra-high dimensional settings infeasible. Moreover, these tests typically suffer from the curse of dimensionality due to the data sparsity in high-dimensional spaces.
Another class of tests for model checking is based on local smoothing techniques, which construct test statistics using the nonparametric estimation of $E[\varepsilon(\beta_0)|X]$ (e.g., \cite{hardle1993}, \cite{zheng1996}, \cite{dette1999}, \cite{fan2001}, \cite{horowitz2001}, \cite{koul2004}, \cite{keilegom2008}). Since these tests rely on nonparametric smoothing, they can typically only detect local alternatives at the nonparametric rate of $O(n^{-1/2}h^{-p/4})$, where $h$ is the bandwidth. Consequently, their statistical properties, including significance level maintenance and power performance, deteriorate rapidly as the dimension $p$ of the predictor vector increases.

There are limited studies in the literature for model checking that are capable of dealing with high dimensional predictors. In diverging dimension settings without sparsity assumption, \cite{tan2019a, tan2022} propose two tests for dense parametric single-index models and multi-index models, respectively. Both of these methods used sufficient dimension reduction and projected empirical processes to address the dimensionality problem. However, they require the predictor dimension $p$ to grow only at a rate of $n^{1/3}$ (up to a logarithmic factor) where $n$ is the sample size, and their direct extension to higher dimensional regimes remains challenging.
In high-dimensional settings with $p>n$, \cite{Shah2018} proposed a residual prediction (RP) test for sparse linear models, utilizing some form of parametric bootstrap to determine critical values. \cite{Jankova2020} extended this idea and developed a generalized residual prediction (GRP) test for sparse generalized linear models. By incorporating data splitting and a debiasing strategy involving the square-root lasso, the GRP test is asymptotically normal, thereby avoiding the bootstrap approximation for the sampling null distribution. It is noted that \cite{Shah2018} mainly focuses on testing high dimensional Gaussian linear models with fixed designs, while \cite{Jankova2020} constructs the test statistic based on projected residuals $w^{\top}\hat{R}$, where $\hat{R}$ is the estimated residual vector and $w$ is a chosen projection in the unit sphere. Since the GRP test focuses on a specific aspect of the model misspecification, it may fail to capture all potential deviations from the null hypothesis and lose power against certain alternatives.

This paper proposes a two-step strategy for testing the goodness-of-fit of sparse regression models under mild assumptions with random designs. Our methodology proceeds as follows. We first construct the projection-based test statistic based on local smoothing techniques. More specifically, the projection-based test statistic is constructed by applying nonparametric estimation to $E[\varepsilon(\beta_0)|\alpha^{\top}X]$, where the projection $\alpha$ lies in the unit sphere of $\mathbb{R}^p$. The asymptotic properties of the projection-based test statistic are established under both the null and alternative hypotheses for any given projection, even when the predictor dimension $p$ grows exponentially with the sample size $n$. An interesting finding is that, for pairwise linearly independent projections, the resulting test statistics are asymptotically independent under the null hypothesis. This theoretical finding
may be of independent interest and could be applied to other statistical methodologies based on projections. Leveraging this property, our second step employs standard combination methods---such as the extreme value combination and the classic Fisher combination \citep{fisher1925}---to aggregate their corresponding $p$-values to form the final test and enhance the statistic power.

A notable feature of our tests is that their asymptotic properties require only the standard convergence rate of parameter estimators, rather than their asymptotic linearity or normality. Consequently, our tests are theoretically applicable to ultra-high dimensional model checking where such asymptotic properties of parameter estimators may fail. Furthermore, the proposed combination tests can be consistent with asymptotic power $1$ under the alternative hypothesis and mild conditions. By projecting the ultra-high dimensional predictor onto a one-dimensional space, our tests can detect local alternatives that depart from the null at the rate of $O(n^{-1/2}h^{-1/4})$. This detection rate is independent of the original dimension $p$ as if the settings were one-dimensional. Therefore, with properly chosen projections, our tests significantly mitigate the curse of dimensionality. The simulation results in Section 5 show the powerfulness of the proposed test compared to existing competitors.

The rest of this paper is organized as follows. Section 2 constructs the projection-based test statistics for any given projections. The limiting null distribution of the projection-based test statistic is established in Section 3. We show in this section that for pairwise linearly independent projections, the resulting test statistics are asymptotically independent under the null. Building on this property, two combinations of their corresponding $p$-values as the final tests are included in this section. This section also provides the power analysis of the proposed tests. Section 4 discusses the selection of projections and provides test statistics for practical use. Simulation studies and two real data analyses to assess the finite sample performance of our tests are presented in Section 5. Section 6 contains discussions and topics for future study. All proofs for the theoretical results are deferred to the Supplementary Material.

\section{Projection-based test statistics}
To illustrate our methodology, we focus our discussion on goodness-of-fit tests for generalized linear models (GLMs). The extension to more general scenarios is straightforward.
Under the GLM framework, the conditional density of $Y$ given $X=x$ is $f(y|x, \beta) = c(y, \sigma) \exp
\{\frac{y\theta(x) - b(\theta(x))}{\sigma^2} \}$,
where $\sigma^2$ is a dispersion parameter, $c(y, \sigma)$ is a positive function and $ \theta(x) = \beta_0^{\top}x $ for some unknown $\beta_0 = (\beta_0^{(1)}, \dots, \beta_0^{(p)})^{\top} \in \mathbb{R}^p$.
It is well known that $E(Y|X=x) = \mu(\beta_0^{\top}x)$ and $var(Y|X=x) = V(\mu(\beta_0^{\top}x))$ for a given inverse link function $\mu(\cdot)$ and a known function $V(\cdot)$. In this paper, we focus on the model checking problem for the conditional mean function $m(x) = E(Y|X=x)$. Thus, the null hypothesis we want to test is
$$ H_0: \mathbb{P}\{ E[Y|X] = \mu(\beta_0^{\top}X) \} = 1, \quad {\rm for \ some } \ \beta_0 \in \Theta \subset \mathbb{R}^p, $$
while the alternative is that the null is false, i.e.,
$$ H_1: \mathbb{P}\{ E[Y|X] \neq \mu(\beta^{\top}X) \} >0, \quad {\rm for \ any } \ \beta \in \Theta \subset \mathbb{R}^p, $$
where $\top$ denotes the transpose and $\Theta$ is a compact set.
In high dimensional settings with $ p \geq n$, as pointed out by \cite{Jankova2020}, if the design matrix ${\bf X} = (X_1, \dots, X_n)^{\top}$ is of full rank, there always exists a solution $\beta_0 \in \mathbb{R}^p$ to the system of linear equations $m(X_i) = \mu(\beta_0^{\top}X_i)$ for $i=1, \dots, n$. This implies that high dimensional GLMs can never be misspecified in practice without any model structural assumption. Following the framework of \cite{Jankova2020}, we consider sparse regression models under both the null and alternative hypotheses. Under the sparsity assumption, the problem of model checking becomes meaningful and well-defined in high dimensional scenarios with $ p \geq n$.

Before introducing our methodology, we present some preliminary notations. For a vector $\beta \in \mathbb{R}^p$, let $\beta^{(j)}$ denote the $j$-th entry of $\beta$ and let $\|\beta\|_q = (\sum_{j=1}^p |\beta^{(j)}|^q)^{1/q} $ for $ q \in \mathbb{Z}^{+}$ and $\|\beta\|_0$ be the number of non-zero entries of $\beta$. For an index set $I \subset \{ 1, 2, \dots, p \}$, $\beta_I$ denotes the vector containing the entries of $\beta$ whose indices are in $I$. For a matrix $A \in \mathbb{R}^{n \times p}$, we define $A_I$ as the matrix containing the columns of $A$ whose indices are in $I$, and $A_{I^c}$ as the matrix containing the columns of $A$ with indices in the complement of $I$. Let $S \subset \{1, \dots, p \}$ be the active set, which represents the indices of $X= (X^{(1)}, \dots, X^{(p)})^{\top}$ that are truly related to the response $Y$. Under the null hypothesis $H_0$, the true regression parameter $\beta_0$ is sparse, and the active set becomes $S=\{j: \beta_0^{(j)} \neq 0 \}$.

Our methodology depends on the following results.
\begin{lemma}\label{lemma-pro}
{\rm (i)} Let $Z \in \mathbb{R}$ and $X \in \mathbb{R}^p$ be random variables. Then we have
\begin{eqnarray*}
E[Z|X] = 0 \ \ a.s.
&\Longleftrightarrow& E[Z|\alpha^{\top}X] =0 \ \ a.s. \quad \forall \ \alpha \in \mathcal{S}^{p-1},
\end{eqnarray*}
where $\mathcal{S}^{p-1} = \{\alpha | \alpha \in \mathbb{R}^p, \|\alpha\|_2 = 1  \}$ is the unit sphere in $\mathbb{R}^p$.

{\rm (ii)} Suppose that $E|Z|^2 < \infty$, $E\| X \|_2^k < \infty$, and $ \sum_{k=1}^{\infty} (E\| X \|_2^k)^{-1/k} = \infty$. If we write
$ \mathcal{A}= \{ \alpha \in \mathbb{R}^{p}: E[Z|\alpha^{\top}X] =0 \ a.s. \} $,  it follows that
\begin{eqnarray*}
E[Z|X] = 0 \ \ a.s. \ \ &\Longleftrightarrow& \ \ \mathcal{A} \ {\rm has \ positive \ Lebesgue \ measure.}
\end{eqnarray*}
\end{lemma}

Part (i) of Lemma~\ref{lemma-pro} follows from Lemma 2.1 of \cite{zhu1998}, Lemma 1 of \cite{escanciano2006a}, or Lemma 2.1 of \cite{lavergne2008}. Part (ii) of Lemma~\ref{lemma-pro} is analogous to Part (B) of Lemma 1 in \cite{patilea2016} and Theorem 2.4 of \cite{cuesta2019}. The condition $\sum_{k=1}^{\infty} (E\| X \|_2^k)^{-1/k} = \infty $ is the so-called Carleman’s condition, which is satisfied if the random vector $X$ has a finite moment generating function in a neighbourhood of the origin, see \cite{cuesta2007} for more details about this condition. The proof of this lemma is provided in the Supplementary Material. Building on Lemma \ref{lemma-pro}, we obtain the following result, which is crucial for the construction of our test statistic.

\begin{corollary}\label{cor-pro}
Suppose that the conditions in part {\rm (ii)} of Lemma \ref{lemma-pro} holds. If we write $ \mathcal{A}_0= \{ \alpha \in \mathcal{S}^{p-1}: E[Z|\alpha^{\top}X] =0, a.s. \} $, then
\begin{eqnarray*}
&& \mathbb{P}\{ E(Z|X)  = 0 \} = 1  \quad  \Longleftrightarrow \quad \mathcal{L}(\mathcal{A}_0) = 1, \\
&& \mathbb{P}\{ E(Z|X) \neq 0 \} >0  \quad \Longleftrightarrow \quad \mathcal{L}(\mathcal{A}_0) = 0,
\end{eqnarray*}
where $\mathcal{L}$ is the uniform probability measure on the unit sphere $\mathcal{S}^{p-1}$.
\end{corollary}

Let $\varepsilon(\beta) = Y-\mu(\beta^{\top}X)$. It is readily seen that the null hypothesis $H_0$ is tantamount to $E[\varepsilon(\beta_0)|X]=0 \ a.s.$ for some $\beta_0 \in \Theta$. For testing $H_0$, according to Corollary~\ref{cor-pro}, we can first choose a suitable projection $\alpha \in \mathcal{S}^{p-1}$ and then for this given projection, test the projected null hypothesis:
$$H_0^{\alpha}: \mathbb{P} \{ E[\varepsilon(\beta_0)|\alpha^{\top}X]=0 \} = 1 \quad {\rm for \ some} \ \beta_0 \in \Theta. $$
The principle of this testing procedure is as follows. Under the null hypothesis $H_0$, the projected null hypothesis $H_0^{\alpha}$ also holds. Conversely, under the alternative $H_1$, we have
$ \mathbb{P} \{E[\varepsilon(\beta_0)|X] \neq 0 \} > 0$. Corollary~\ref{cor-pro} implies the projected null hypothesis $H_0^{\alpha}$ fails for $\mathcal{L}$-almost every $\alpha \in \mathcal{S}^{p-1}$ under the alternative $H_1$.
Consequently, the null hypothesis $H_0$ is $\mathcal{L}$-a.s. equivalent to the projected null hypothesis $H_0^{\alpha}$.

According to Lemma \ref{lemma-pro} and Corollary \ref{cor-pro}, we construct the test statistics based on the projected null hypothesis $ H_0^{\alpha}: E[\varepsilon(\beta_0)|\alpha^{\top}X]=0 \ a.s.$ for some $\beta_0 \in \Theta$.
For a given projection $\alpha \in \mathcal{S}^{p-1}$, $H_0^{\alpha}$ is equivalent to
\begin{align}\label{popu-process}
E\{\varepsilon(\beta_0) E[\varepsilon(\beta_0)|\alpha^{\top}X]f_{\alpha}(\alpha^{\top}X)\} = E\{ E[\varepsilon(\beta_0) |\alpha^{\top}X]^2f_{\alpha}(\alpha^{\top}X)\} = 0,
\end{align}
for some $\beta_0 \in \Theta \subset \mathbb{R}^p$, where $f_{\alpha}(\cdot)$ denotes the density function of $\alpha^{\top}X$.
Let $(X_i, Y_i)_{i=1}^n $ be an i.i.d. sample with the same distribution as $(X, Y) \in \mathbb{R}^{p+1}$ and let $\hat{\beta}_0$ be an estimator of $\beta_0$ under the GLM setting, such as a penalized estimator or its variants.
We then construct the (informal) projection-based test statistic for checking $H_0^{\alpha}$ based on (\ref{popu-process}) as
\begin{align*}
\hat{V}_n^{\alpha} = \frac{1}{n(n-1)} \sum_{i \neq j}^n \varepsilon_i(\hat{\beta}_0) \varepsilon_j(\hat{\beta}_0) K_h(\alpha^{\top}X_i - \alpha^{\top}X_j),
\end{align*}
where $\varepsilon_i(\hat{\beta}_0) = Y_i - \mu(\hat{\beta}_0^{\top}X_i)$, and $K_h(\cdot) = h^{-1} K(\cdot/h)$ with $K(\cdot)$ being a univariate kernel function and $h$ being the bandwidth.

\begin{remark}
It is worth mentioning that \cite{cuesta2019} utilized a similar idea for testing the goodness-of-fit of functional linear models. Their test statistic is based on projected empirical processes and requires the asymptotic normality of parameter estimators to derive the limiting null distribution.
In contrast, as we show in the next section, the asymptotic properties of the projection-based test $\hat{V}_n^{\alpha}$ requires only a convergence rate of the parameter estimators, rather than their asymptotic normality.
\end{remark}

\section{Main results}
\subsection{Limiting null distribution of projection-based test statistics}
In this subsection, we investigate the asymptotic properties of $\hat{V}_n^{\alpha}$ under the null $H_0$ when the covariate dimension $p$ may be much larger than the sample size $n$. For this, we further introduce some notion. A random variable $Z \in \mathbb{R}$ is called sub-Weibull of order $\tau > 0$, if
$$ \|Z\|_{\psi_{\tau}}:=\inf\{\eta>0: E[\psi_{\tau}(\frac{|Z|}{\eta})] \leq 1 \} < \infty,  $$
where $\psi_{\tau}(x) = \exp(x^{\tau}) - 1$ for $x \geq 0$.
By the Markov inequality, if $Z$ is sub-Weibull of order $\tau$, then
$ \mathbb{P}(|Z| > t) \leq 2 \exp(-t^{\tau}/\|Z\|_{\psi_{\tau}}^{\tau} )$
for all $t \geq 0$.
It is readily seen that sub-Gaussian and sub-exponential random variables are the special cases of sub-Weibull distributions with $\tau=2$ and $\tau=1$, respectively. Note that sub-Weibull variables are not required to be zero-mean. More details about sub-Weibull random variables can be found in \cite{Vladimirova2020} and \cite{Kuchibhotla2022}. The following conditions are needed to derive the asymptotic properties of $\hat{V}_n^{\alpha}$ under the null $H_0$.
Let $\hat{S} = \{j: \hat{\beta}_0^{(j)} \neq 0\}$, $\hat{s} = |\hat{S}|$, and $s=|S|$.
The notation $C$ throughout this paper denotes a generic constant independent of $n$, which may differ across appearances.

(A1) Under the null $H_0$, the estimator $\hat{\beta}_0$ satisfies $\|\hat{\beta}_0 - \beta_0 \|_2 = O_p(\sqrt{\frac{s \log{p}}{n}})$ and $\| \hat{\beta}_0 - \beta_0 \|_1 = O_p(\sqrt{\frac{s^2 \log{p}}{n}})$.

(A2) The random variables $\varepsilon(\beta_0)$, $\mu'(\beta_0^{\top}X)$, and $\mu''(\beta_0^{\top}X)$ are sub-Weibull of order $\tau =2$ with $\max\{
\| \varepsilon(\beta_0) \|_{\psi_2}, \| \mu'(\beta_0^{\top}X) \|_{\psi_2}, \|\mu''(\beta_0^{\top}X) \|_{\psi_2}\} \leq C < \infty $.
The covariates $X \in \mathbb{R}^p$ is centered and $\alpha^{\top}X$ is sub-Weibull of order $\tau =2$ with $ \| \alpha^{\top}X \|_{\psi_2} \leq C < \infty $ for all $\alpha \in \mathcal{S}^{p-1}$.

(A3) The link function $\mu(\cdot)$ admits third derivatives, and $|\mu'''(\beta^{\top}x)| \leq F(x) $ for all $\beta \in \Theta$ with $F(X)$ being sub-Weibull of order $\tau \geq 1/3$.

(A4) The univariate kernel function $K(\cdot)$ is a continuous, bounded, and symmetric function such that $\int K(u) du =1$, $\int |u|K^2(u) du < \infty$, $\int |u|K^4(u) du < \infty$, and $\int |u+v|^i K^2(u+v) dudv < \infty$ for $i =1, 2, 3$.

(A5) We write $\sigma_{\alpha}^2(t) = E[\varepsilon^2(\beta_0)| \alpha^{\top}X = t]$, $\kappa_{\alpha}^4(t) = E[\varepsilon^4(\beta_0) | \alpha^{\top}X = t]$, and $g_{\alpha, \theta }(t) = E[\theta^{\top}X \mu'(\beta_0^{\top}X)|\alpha^{\top}X = t] $. The functions $\sigma^2_{\alpha}(t)$, $\kappa^2_{\alpha}(t)$, and $g_{\alpha, \theta }(t)$ satisfy that for any $u \in \mathbb{R}$, $\alpha, \theta \in \mathcal{S}^{p-1}$,
\begin{eqnarray*}
|\sigma^2_{\alpha}(t+u)f_{\alpha}(t+u) - \sigma^2_{\alpha}(t)f_{\alpha}(t)| &\leq& L|u|, \\
|\kappa^4_{\alpha}(t+u)f_{\alpha}(t+u) - \kappa^4_{\alpha}(t)f_{\alpha}(t)| &\leq& L|u|, \\
|g_{\alpha, \theta}(t+u)f_{\alpha}(t+u) - g_{\alpha, \theta}(t)f_{\alpha}(t)| &\leq& L|u|,
\end{eqnarray*}
where $f_{\alpha}(t)$ denotes the density function of $\alpha^{\top}X$.

(A6) The asymptotic variance $\Sigma^{\alpha}$ of $nh^{1/2} \hat{V}_n^{\alpha}$ under the null $H_0$ satisfies that $\Sigma^{\alpha} \geq C >0$ for all $\alpha \in \mathcal{S}^{p-1}$.

Condition (A1) is satisfied by many popular sparse estimators and their variants under certain regularity conditions, such as the GLM Lasso estimator and the GLM SCAD estimator, when the underlying GLMs are correctly specified. In the Supplementary Material, we present the detailed conditions such that (A1) is satisfied for Lasso estimators.
The assumption of sub-Weibull of order $2$ (sub-Gaussian) in (A2) is usually imposed in the literature of high dimensional data analysis, see \cite{buhlmann2011} for instance. It is used to bound the tail probability of $U$-statistics in the decomposition of $\hat{V}_n^{\alpha}$ when the dimension $p$ may significantly exceed the sample size $n$.
Condition (A3) is satisfied by many GLMs that are usually used in practice, such as Gaussian linear models, logistic regression models, and probit regression models. It is used to control the convergence rate of the remainder term in the decomposition of $\hat{V}_n^{\alpha}$.
The sub-Weibull order $\tau \geq 1/3$ of $F(X)$ in (A3) is just a technical condition that can be weakened if we impose a more restrictive condition on the divergence rate of $p$ in Theorem \ref{th-Vn}.
Conditions (A4) and (A5) are commonly used in the literature on nonparametric estimations, see \cite{rao1983} and \cite{zhu1996} for instance.
Condition (A6) is used to ensure the weak convergence of the projection-based test statistic $\hat{T}_n^{\alpha}$ defined in (\ref{Tn-alpha}).

The next theorem establishes the asymptotic property of $ \hat{V}_n^{\alpha}$ under the null hypothesis in ultra-high dimensional settings. 

\begin{theorem}\label{th-Vn}
Suppose that Conditions (A1)-(A6) hold. If $s h^{1/2} \log{p} = o(1)$, $s^4 \log^3{p} =o(nh)$, and $\log^5{p}=o(n)$ as $n \to \infty$, then under the null $H_0$ and for a given projection $\alpha \in \mathcal{S}^{p-1}$,
$$ nh^{1/2} \hat{V}_n^{\alpha}/(\Sigma^{\alpha})^{1/2} \longrightarrow N(0, 1), \quad  in \ distribution, $$
where $\Sigma^{\alpha} = 2\int \sigma_{\alpha}^4(t) f_{\alpha}^2(t)dt \int K^2(u)du $ is the asymptotic variance of $nh^{1/2} \hat{V}_n^{\alpha}$, which can be estimated by
$$ \hat{\Sigma}^{\alpha} = \frac{2h}{n(n-1)}\sum_{i \neq j} \varepsilon_i^2(\hat{\beta}_0) \varepsilon_j^2(\hat{\beta}_0) K_h^{2}(\alpha^{\top}X_{i}- \alpha^{\top}X_{j}). $$
\end{theorem}

The proof of Theorem~\ref{th-Vn} is provided in the Supplementary Material. According to Theorem~\ref{th-Vn}, our final projection-based test statistic is
\begin{equation}\label{Tn-alpha}
\hat{T}_n^{\alpha}
= \sqrt{\frac{n-1}{n}} \frac{nh^{1/2} \hat{V}_n^{\alpha}}{(\hat{\Sigma}^{\alpha})^{1/2}}
= \frac{\sum_{1\leq i \neq j \leq n} \varepsilon_i(\hat{\beta}_0)
  \varepsilon_j(\hat{\beta}_0)K(\frac{\alpha^{\top}X_i - \alpha^{\top}X_j}{h})}{\left(2 \sum_{1\leq i \neq j \leq n} \varepsilon_i^2(\hat{\beta}_0) \varepsilon_j^2(\hat{\beta}_0) K^2(\frac{\alpha^{\top}X_i - \alpha^{\top}X_j}{h}) \right)^{1/2}}.
\end{equation}
The limiting null distribution of $\hat{T}_n^{\alpha}$ follows from Theorem~\ref{th-Vn}.

\begin{corollary}\label{cor-Tn}
Assume the conditions in Theorem~\ref{th-Vn}. Then under the null $H_0$ and for a given projection $\alpha \in \mathcal{S}^{p-1}$, we have
$$ \hat{\Sigma}^{\alpha} = \Sigma^{\alpha} + o_p(1), $$
and consequently,
$$ \hat{T}_n^{\alpha} \longrightarrow N(0, 1), \quad in \ distribution. $$
\end{corollary}

\subsection{Combined projection-based tests}
With a set of chosen projections $\alpha \in \mathcal{S}^{p-1}$, we obtain a series of projection-based test statistics $\hat{T}_n^{\alpha}$. Although the test $\hat{T}_n^{\alpha}$ is $\mathcal{L}$-a.s. consistent under the alternative hypothesis, it may still lose power for certain specific projections. Another potential problem is that the values of the test statistics $\hat{T}_n^{\alpha}$ may vary across different projections, which may also affect the power performance of the projection-based tests. To address these limitations, a natural idea is to combine the various projection-based test statistics $\hat{T}_n^{\alpha}$ to form a final test statistic to enhance power.

An interesting result is that for pairwise linearly independent projections $\alpha_1, \dots, \alpha_d \in \mathcal{S}^{p-1}$, the resulting test statistics $ \hat{T}_n^{\alpha_1}, \dots, \hat{T}_n^{\alpha_d}$ are asymptotically independent under the null hypothesis. Here, pairwise linear independence means that $k_1 \alpha_i + k_2 \alpha_j = 0$ if and only if $k_1=k_2=0$ for all $1 \leq i \neq j \leq d$.
For illustration, let $d=2$. It follows from the proofs of Theorem \ref{th-Vn} and Corollary~\ref{cor-Tn} that under $H_0$,
$$ \hat{T}_n^{\alpha} = \frac{nh^{1/2} V_{n0}^{\alpha}}{(\Sigma^{\alpha})^{1/2}} + o_p(1), $$
where $ V_{n0}^{\alpha} = \frac{2}{n(n-1)}\sum_{1 \leq i < j \leq n} \varepsilon_i(\beta_0) \varepsilon_j(\beta_0) K_h(\alpha^{\top}X_i - \alpha^{\top}X_j)$ and $\Sigma^{\alpha} = 2\int \sigma_{\alpha}^4(t) f_{\alpha}^2(t)dt \int K^2(u)du $.
Next, we show that the covariance of $(nh^{1/2} V_{n0}^{\alpha_1}, nh^{1/2} V_{n0}^{\alpha_2})$ converges to zero when $\alpha_1, \alpha_2 \in \mathcal{S}^{p-1}$ are linearly independent. To this end, we impose an additional condition.

(A7) Let $\sigma_{\alpha_{12}}^2(s, t) = E(\varepsilon^2(\beta_0)|\alpha_1^{\top}X=s, \alpha_2^{\top}X=t)$ and $f_{\alpha_{12}}(s,t)$ be the joint density function of $(\alpha_1^{\top}X, \alpha_2^{\top}X)$. The function $\sigma_{\alpha_{12}}^2(s, t) f_{\alpha_{12}}(s,t)$ satisfies that for any $u, v \in \mathbb{R}$,
$$ |\sigma_{\alpha_{12}}^2(s+u, t+v) f_{\alpha_{12}}(s+u,t+v) - \sigma_{\alpha_{12}}^2(s, t) f_{\alpha_{12}}(s,t) | \leq L(|u|+|v|).$$

By Condition (A7) and some elementary calculations, we can show that
$$ cov(nh^{1/2} V_{n0}^{\alpha_1}, nh^{1/2} V_{n0}^{\alpha_2}) = 2h \int \sigma_{\alpha_{12}}^4(s, t)f_{\alpha_{12}}^2(s,t)dsdt + O(h^2) = o(1). $$
Note that under $H_0$,
$$ (\hat{T}_n^{\alpha_1}, \hat{T}_n^{\alpha_2}) = \left( \frac{nh^{1/2} V_{n0}^{\alpha_1}}{(\Sigma^{\alpha_1})^{1/2}}, \frac{nh^{1/2} V_{n0}^{\alpha_2}}{(\Sigma^{\alpha_2})^{1/2}} \right) + o_p(1).  $$
Under certain regularity conditions, we can show that $(\hat{T}_n^{\alpha_1}, \hat{T}_n^{\alpha_2})$ are asymptotically independent under $H_0$. A more specific result is summarized in the following theorem, and its proof is provided in the Supplementary Material.

\begin{theorem}\label{multi-Tn}
Suppose that (A7) and the conditions in Theorem~\ref{th-Vn} hold, and the projections $\alpha_1, \dots, \alpha_d \in \mathcal{S}^{p-1}$ are pairwise linearly independent. Then under $H_0$, we have
$$ (\hat{T}_n^{\alpha_1}, \dots, \hat{T}_n^{\alpha_d}) \longrightarrow N(0, I_d), \quad in \ distribution, $$
where $N(0, I_d)$ denotes the $d$-dimension standard normal distribution.
\end{theorem}

For each projection $\alpha_i$, the asymptotic $p$-value of $\hat{T}_n^{\alpha_i}$ is given by
$ \hat{p}_{\alpha_i} = 1-\Phi(\hat{T}_n^{\alpha_i})$,
where $\Phi(\cdot)$ denotes the standard normal cumulative distribution function.
By Theorem \ref{multi-Tn} and the Continuous mapping theorem, we have under $H_0$,
\begin{eqnarray}\label{multi-p}
(\hat{p}_{\alpha_1}, \dots, \hat{p}_{\alpha_d})  \longrightarrow (p_1, \dots, p_d), \quad {\rm in \ distribution},
\end{eqnarray}
where $p_1, \dots, p_d$ are i.i.d. uniform random variables on $(0,1)$.
Therefore, we can construct combined test statistics using various powerful combinations of these $p$-values, such as the minimum of $p$-values or the classic Fisher’s combination method \citep{fisher1925}.
The combined test statistics based on the minimum of $p$-values and the Fisher combination are defined as
\begin{eqnarray}\label{integrated-stat}
\hat{T}_{\text{min}} = \min\{\hat{p}_{\alpha_1}, \dots, \hat{p}_{\alpha_d}\},  \ \ \text{and} \ \
\hat{T}_{\text{Fisher}} = -2 \sum_{i=1}^d \ln(\hat{p}_{\alpha_i}),
\end{eqnarray}
respectively. The following result presents the limiting null distribution of $\hat{T}_{\text{min}}$ and $\hat{T}_{\text{Fisher}}$, which is a consequence of (\ref{multi-p}) and the Continuous Mapping Theorem.

\begin{corollary}\label{limit_null_integ_sta}
Assume the conditions in Theorem~\ref{multi-Tn}. \\
(1) The combined test statistic $\hat{T}_{\rm{min}}$ satisfies that, under $H_0$,
\begin{eqnarray*}
\hat{T}_{\rm{min}}  \longrightarrow p_{(1)}, \quad in \ distribution,
\end{eqnarray*}
where $p_{(1)} = \min\{p_1, \dots, p_d\}$ has a cumulative distribution function $F_{p_{(1)}}(u) = [1-(1-u)^d]I(0 \leq u \leq 1)$. \\
(2) The combined test statistic $ \hat{T}_{\rm{Fisher}} $ satisfies that, under $H_0$,
\begin{eqnarray*}
\hat{T}_{\rm{Fisher}} \longrightarrow \chi^2(2d),  \quad in \ distribution,
\end{eqnarray*}
where $\chi^2(2d)$ denotes a chi-square distribution with $2d$ degree of freedom.
\end{corollary}

It follows from Corollary~\ref{limit_null_integ_sta} that the asymptotic critical values of the combined tests $ \hat{T}_{\text{min}} $ and $\hat{T}_{\text{Fisher}}$ can be determined by quantiles of the distributions of $p_{(1)}$ and $\chi^2(2d)$, respectively. This implies that both $ \hat{T}_{\text{min}} $ and $\hat{T}_{Fisher}$ are asymptotically distribution-free. Consequently, we do not need to resort to resampling methods, such as the wild bootstrap, to approximate the limiting null distribution.
It is worth noting that selecting appropriate projections is a non-trivial task. An unsuitable choice can lead to a significant loss of power under alternative hypotheses, especially in ultra-high dimensional scenarios. Section~5 provides a detailed discussion on how to select these projections to ensure that the proposed tests maintain good power performance.

\begin{remark}\label{dim-curse}
Note that the projection-based test statistic $\hat{T}_n^{\alpha}$ only relies on one-dimensional nonparametric estimation. Thus, by choosing appropriate projections, our tests can effectively mitigate the curse of dimensionality, even when the predictor dimension $p$ substantially exceeds the sample size $n$. \cite{lavergne2008} proposed a dimension-reduction approach for testing parametric regression models using a statistic based on $\hat{V}_n^{\alpha}$ in fixed dimensional settings. They selected the projection by maximizing $nh^{1/2} \hat{V}_n^{\alpha}$ plus a penalty over all $\alpha \in \mathcal{S}^{p-1}$. As a result, $\hat{V}_n^{\alpha}$ is treated as a $U$-process indexed by $\alpha \in \mathcal{S}^{p-1}$ \citep[Lemma 3.1]{lavergne2008}. However, in high dimensional settings with $p \geq n$, it becomes challenging to derive the asymptotic properties of the $U$-process $\hat{V}_n^{\alpha}$, making it difficult to extend their methodology to high dimensional cases.
\end{remark}

\begin{remark}\label{irrepresent-cond}
It is also worth noting that we do not require the variable selection consistency of the estimator $\hat{\beta}_0$ in high dimensional settings---that is, $\mathbb{P}(\hat{S} = S) \to 1$ as $n \to \infty$---in Theorem \ref{th-Vn} and Corollary \ref{cor-Tn}.
If this consistency holds for $\hat{\beta}_0$, then we only need to work on the event $\{ \hat{S} = S \}$ and the proofs for Theorem \ref{th-Vn} and Corollary \ref{cor-Tn} can be substantially simplified.
However, the variable selection consistency for Lasso estimators typically requires rather restrictive conditions, such as Irrepresentable Condition \citep{Meinshausen2006, zhao2006, zou2006}. By exploring the tail probabilities of the sums of sub-Weibull random variables and $U$-statistics, we find that our proofs for Theorem \ref{th-Vn} and Corollary \ref{cor-Tn} do not need the variable selection consistency of a penalized estimator $\hat{\beta}_0$. This implies that, in theory, our tests are applicable to any estimator satisfying Condition (A1).
Furthermore, the growing rate of $p$ in Theorem \ref{th-Vn} is permitted to be of exponential order relative to the sample size $n$, when the bandwidth $h$ is properly chosen. This property highlights the applicability of our tests for assessing the goodness of fit of ultra-high dimensional regression models.
\end{remark}

\subsection{Power analysis}
We now investigate the power analysis of the proposed tests under the local and global alternative hypotheses. Consider a sequence of alternative hypotheses that converge to the null $H_0$:
\begin{equation}\label{alter-hypo}
 H_{1n}: m(x) = E(Y|X=x) = \mu(\beta_0^{\top}x) + r_n R(x),
\end{equation}
where $r_n $ is a sequence of positive numbers converging to $0$ as $n \to \infty$, and $R(\cdot)$ is a non-constant function satisfying $\mathbb{P}(R(X) = 0) < 1$. We also assume a sparsity structure for the regression function $m(\cdot)$.
To derive the asymptotic properties of $\hat{T}_n^{\alpha}$ under the alternative hypotheses, we impose the following additional conditions.

(A8). (i) Suppose that under the global alternative $H_1$, there exists $\tilde{\beta}_0$ in the interior of $\Theta$ satisfying
$$ \|\hat{\beta}_0 - \tilde{\beta}_0\|_1 = O_p(\tilde{s} \sqrt{\frac{\log{p}}{n}})
\quad {\rm and} \quad
\|\hat{\beta}_0 - \tilde{\beta}_0\|_2 = O_p( \sqrt{\frac{\tilde{s} \log{p}}{n}}), $$
where $\tilde{s} = |\tilde{S}|$ with $\tilde{S}$ being the support of $\tilde{\beta}_0$.
(ii) Under the local alternative $H_{1n}$ with $r_n \to 0$, we have $\tilde{\beta}_{0} - \beta_{0} = r_n M^L + o_p(r_n)$ with $M^L_{\tilde{S}_1^c} = 0$ and $\| M^L \|_2 = O(1)$,
$$ \|\hat{\beta}_0 - \beta_0\|_1 = O_p(\tilde{s} \sqrt{\frac{\log{p}}{n}} + \sqrt{\tilde{s}_1} r_n), \quad {\rm and} \quad  \|\hat{\beta}_0 - \beta_0\|_2 = O_p( \sqrt{\frac{\tilde{s} \log{p}}{n}} + r_n),$$
where $\tilde{s}_1 = |\tilde{S}_1|$ and $\tilde{S}_1 = S \cup \tilde{S}$ with $S$ being the support of $\beta_0$ in (\ref{alter-hypo}).

(A9) The random variables $\varepsilon(\tilde{\beta}_0)$, $\mu'(\tilde{\beta}_0^{\top}X)$, and $\mu''(\tilde{\beta}_0^{\top}X)$ are sub-Weibull of order $\tau =2$ with $\max\{
\| \varepsilon(\tilde{\beta}_0) \|_{\psi_2}, \| \mu'(\tilde{\beta}_0^{\top}X) \|_{\psi_2}, \|\mu''(\tilde{\beta}_0^{\top}X) \|_{\psi_2}\} \leq C < \infty $.

(A10) Let $\tilde{\sigma}_{\alpha}^2(t) = E[\varepsilon^2(\tilde{\beta}_0) |\alpha^{\top}X = t]$, $\tilde{g}_{\alpha}(t) = E[\varepsilon(\tilde{\beta}_0)|\alpha^{\top}X = t]$, $\tilde{g}_{\alpha, \theta }(t) = E[\theta^{\top}X \mu' (\tilde {\beta}_0^{\top} X)|\alpha^{\top}X=t] $, and $ R_{\alpha}^{(i)}(t) = E[R^i(X)|\alpha^{\top}X = t]$ with $i=1, 2$.
The functions $\tilde{\sigma}_{\alpha}^2(t)$, $\tilde{g}_{\alpha}(t)$, $\tilde{g}_{\alpha, \theta }(t)$, and $R^{(i)}_{\alpha}(t)$ satisfy that for all $u \in \mathbb{R}$, $\alpha, \theta \in \mathbb{S}^{p-1}$,
\begin{eqnarray*}
|\tilde{\sigma}^2_{\alpha}(t+u)f_{\alpha}(t+u) - \tilde{\sigma}^2_{\alpha}(t)f_{\alpha}(t)| &\leq& L|u|, \\
|\tilde{g}_{\alpha}(t+u)f_{\alpha}(t+u)-\tilde{g}_{\alpha}(t)f_{\alpha}(t)|
&\leq& L|u| \\
|\tilde{g}_{\alpha, \theta }(t+u)f_{\alpha}(t+u) - \tilde{g}_{\alpha, \theta }(t)f_{\alpha}(t)|
&\leq& L|u| \\
|R^{(i)}_{\alpha}(t+u)f_{\alpha}(t+u) - R^{(i)}_{\alpha}(t)f_{\alpha}(t)| &\leq& L|u|.
\end{eqnarray*}

In the Supplementary Material, we show that the GLM Lasso estimator satisfies Condition (A8) under both the global and local alternative hypotheses. The asymptotic properties of other penalized estimators or their variants can be established similarly under the alternatives. \cite{lu2012} and \cite{Buhlmann2015} showed that the Lasso and adaptive Lasso estimators, respectively, satisfy Part ($\mathrm{i}$) of Condition (A8). They further showed that the support $\tilde{S}$ of $\tilde{\beta}_0$ can be a subset of the true active set $S$ under the alternative hypothesis (i.e., the misspecified model). Their results also provide justification for Condition (A8). Conditions (A9) and (A10) are similar to (A2) and (A5), respectively, under the null $H_0$.
They are used to bound the convergence rate of the remainders in the decomposition of $\hat{V}_n^{\alpha}$ under the alternatives.

The next theorem establishes the asymptotic properties of the projection-based test statistic $ \hat{T}_n^{\alpha} $ for a given $\alpha$ under various alternative hypotheses in ultra-high dimensional settings. Its proof is provided in the Supplementary Material. We write
\begin{align*}
\tilde{V}^{\alpha} &= E\{ [E(\varepsilon (\tilde{\beta}_0)|\alpha^{\top}X)]^2 f_{\alpha}(\alpha^{\top}X) \}, \\
\tilde{\Sigma}^{\alpha} &= 2\int \tilde{\sigma}_{\alpha}^4(t) f_{\alpha}^2(t) dt \int K^2(u)du, \\ \gamma_n^{\alpha} &= E\{ |E[M_L^{\top}X \mu'(\beta_0^{\top}X) + R(X)|\alpha^{\top}X] |^2 f_{\alpha}(\alpha^{\top}X) \}.
\end{align*}

\begin{theorem}\label{th-power}
Suppose that Conditions (A2)-(A6) and (A8)-(A10) hold. \\
(1) Under the global alternative $H_1$, if $ \tilde{s}^2 \log{p} = o(nh) $ and $\log{p} = o(n^{1/5})$ as $n \to \infty$, then
\begin{align*}
\hat{V}_{n}^{\alpha}  = \tilde{V}^{\alpha} + o_p(1), \quad {\rm and} \quad
\hat{\Sigma}^{\alpha} = \tilde{\Sigma}^{\alpha} + o_p(1).
\end{align*}
(2) Under the local alternative $H_{1n}$ with $r_n = 1/\sqrt{nh^{1/2}}$, if $\tilde{s}_1 h^{1/2} \log{p} = o(1)$ and $\tilde{s}_1 = o(n h^{5/2})$ as $n \to \infty$, then
\begin{eqnarray*}
nh^{1/2} \hat{V}_{n}^{\alpha}/\hat{\Sigma}^{\alpha} \longrightarrow N(\gamma, 1), \quad in \ distribution.
\end{eqnarray*}
where $\gamma$ is the limit of $\gamma_n^{\alpha}/\Sigma^{\alpha}$ and $\Sigma^{\alpha}=2\int \sigma_{\alpha}^4(t) f_{\alpha}^2(t)dt \int K^2(u)du$ is given in Theorem~\ref{th-Vn}.
\end{theorem}

Recall that the projection-based test statistic is defined as $\hat{T}_n^{\alpha} = \sqrt{n(n-1)h} \hat{V}_{n}^{\alpha} /\hat{\Sigma}^{\alpha}$. Under the alternative $H_1$, it follows from Corollary \ref{cor-pro} that $ \mathbb{P}\{ E[\varepsilon(\tilde{\beta}_0)|\alpha^{\top}X] \neq 0 \} >0 $ and then
$ \tilde{V}^{\alpha} > 0$ for almost every $\alpha \in \mathcal{S}^{p-1}$. If $\tilde{\Sigma}^{\alpha}$ is bounded and $\tilde{V}^{\alpha}$ is bounded away from zero, it follows from Theorem \ref{th-power} that $\hat{T}_n^{\alpha}$ diverges to infinity at the rate $O(nh^{1/2})$ under $H_1$. This implies that the projection-based test $\hat{T}_n^{\alpha}$ is consistent with the asymptotic power 1 under $H_1$ in ultra-high dimensional settings.
Furthermore, if $\gamma \neq 0$, it follows from Theorems~\ref{th-Vn} and~\ref{th-power} that the test $\hat{T}_n^{\alpha}$ can detect local alternatives distinct from the null at the rate of $O(n^{-1/2}h^{-1/4})$.

Combining Theorem~\ref{th-power} with the Slutsky theorem and Continuous mapping theorem, we readily obtain the asymptotic properties of the combination tests $\hat{T}_{\min}$ and $\hat{T}_{\rm{Fisher}}$ under the alternatives.

\begin{corollary}\label{limit-alter-minFisher}
Suppose that Conditions (A2)-(A6) and (A8)-(A10) hold. \\
(1) Under the global alternative $H_1$, if $\tilde{\Sigma}^{\alpha_i} \leq C < \infty$ and $\tilde{V}^{\alpha_i} \geq C >0$ for some projection $\alpha_i \in \mathcal{S}^{p-1}$, $ \tilde{s}^2 \log{p} = o(nh) $ and $\log{p} = o(n^{1/5})$ as $n \to \infty$, then, in probability,
$$ \hat{T}_{\min} \longrightarrow 0 \quad {\rm and} \quad \hat{T}_{\rm{Fisher}} \longrightarrow \infty.  $$
(2) Under the local alternative $H_{1n}$ with $r_n = n^{-1/2}h^{-1/4}$, if $\tilde{s}_1 h^{1/2} \log{p} = o(1)$ and $\tilde{s}_1 = o(n h^{5/2})$ as $n \to \infty$, then, in distribution,
\begin{align*}
\hat{T}_{\min} \longrightarrow \tilde{p}_{(1)},
\quad {\rm and } \quad
\hat{T}_{\rm{Fisher}} \longrightarrow -2 \sum_{i=1}^d \ln(\tilde{p}_{\alpha_i}),
\end{align*}
where $\tilde{p}_{(1)} = \min\{ \tilde{p}_1, \dots, \tilde{p}_d\}$, $\tilde{p}_i = 1- \Phi(\tilde{T}^{i})$, and $(\tilde{T}^1, \dots, \tilde{T}^d)^{\top} \sim N(\tilde{\gamma}, I_d)$ with $\tilde{\gamma} \in \mathbb{R}^d$ being the limit of $(\gamma_n^{\alpha_1}/\Sigma^{\alpha_1}, \dots, \gamma_n^{\alpha_d}/\Sigma^{\alpha_d})^{\top}$.
\end{corollary}

It follows from Theorem~\ref{th-power} and Corollary~\ref{limit-alter-minFisher} that under some regularity conditions, the combination tests $\hat{T}_{\min}$ and $\hat{T}_{\rm{Fisher}}$ can be
consistent with the asymptotic power $1$ under $H_1$ in ultra-high dimensional settings. Furthermore, Corollary~\ref{limit-alter-minFisher} shows that our tests $\hat{T}_{\min}$ and $\hat{T}_{\rm{Fisher}}$
can detect local alternatives distinct from the null at the rate of $O(n^{-1/2}h^{-1/4})$, even if the dimension $p$ substantially exceeds the sample size $n$.
This detection rate is in line with the results of dimension-reduction tests proposed by \cite{lavergne2008, lavergne2012} in fixed dimensional settings. In contrast, classic local smoothing tests \citep{hardle1993, zheng1996} typically can only detect local alternatives that converge to the null at the rate of $ O( n^{-1/2}h^{-p/4}) $. Thus, these tests suffer severely from the curse of dimensionality and cannot be applied in ultra-high dimensional scenarios. Note that our tests detect the alternative as if the predictor were one-dimensional. Since the detection rate is independent of the predictor dimension, this suggests that our tests, when combined with suitable projections, can effectively mitigate the curse of dimensionality even in ultra-high dimensional settings.

\section{Test statistics for practical use}
In practice, we must delicately choose the projections since the power performance of our proposed tests is obviously affected by projections. We illustrate this issue in the context of testing high-dimensional Gaussian linear models. Without loss of generality, we assume the data are standardized such that $E(X) =0$ and $E(Y)=0$. Note that the test statistic $\hat{T}_n^{\alpha}$ has power under the alternative hypothesis only if $E\{ [E(\varepsilon(\tilde{\beta}_0) |\alpha^{\top}X)]^2 f_{\alpha}(\alpha^{\top}X) \} > 0 $ for the projection $\alpha \in \mathcal{S}^{p-1}$, where $\varepsilon(\tilde{\beta}_0) = Y- \tilde{\beta}_0^{\top}X $. Consider the extreme case where $\varepsilon(\tilde{\beta}_0) $ is mean independent of $\alpha^{\top}X$ for a specific projection $\alpha$; that is, $E[\varepsilon(\tilde{\beta}_0)|\alpha^{\top}X] = E[\varepsilon(\tilde{\beta}_0)]$. We provide a toy example to show that this situation can arise in high dimensional scenarios. Let $Y = m(X) + \varepsilon$ and $S$ be the true active set. Suppose that $X \sim N(0, I_p)$ and the error term $\varepsilon$ is independent of $X$. Under certain regularity conditions, \cite{Buhlmann2015} showed that the support $\tilde{S}$ of $\tilde{\beta}_0$ satisfies $\tilde{S} \subset S$. Let $S_{\alpha}$ be the support of the projection $\alpha$. If $ S_{\alpha} \subset S^c $, then $ \varepsilon(\tilde{\beta}_0) = m(X_S) - \tilde{\beta}_{0S}^{\top}X_{S} + \varepsilon$ is independent of $\alpha^{\top}X$ and thus $E[\varepsilon(\tilde{\beta}_0)|\alpha^{\top}X] = E[\varepsilon(\tilde{\beta}_0)]$. In this case, it is readily seen that $E[\varepsilon(\tilde{\beta}_0)|\alpha^{\top}X] =0$ and thus $E\{ [E(\varepsilon(\tilde{\beta}_0) |\alpha^{\top}X)]^2 f_{\alpha}(\alpha^{\top}X)\} = 0 $. This implies that the test $\hat{T}_n^{\alpha}$ may have no power under the alternatives in such a situation.
To ensure our test has high power, we choose the projections such that the projected predictors $\alpha^{\top}X$ are strongly associated with the residual $\varepsilon(\tilde{\beta}_0)$. We therefore propose a data-driven method for selecting these projections. The principle of our method is as follows.

We assume, without loss of generality, that $\varepsilon(\tilde{\beta}_0) = g(\tilde{\vartheta}, \tilde{\beta}_1^{\top}X, \dots, \tilde{\beta}_l^{\top}X, \varepsilon)$ follows a multiple-index model where $\tilde{\vartheta} \in \mathbb{R}$, $\tilde{\beta}_1, \dots, \tilde{\beta}_l \in \mathcal{S}^{p-1} $ are the latent projections, and $\varepsilon$ is an error term. If there were no dimension reduction structure in this model, then we would have $l = p$ and $\tilde{\beta}_i = (0, \dots, 0, 1, 0, \dots, 0)^{\top}$ with $1$ in the $i$-th component and $0$ otherwise. However, this scenario may not arise as we focus on sparse regression models under both the null and alternative hypotheses. Furthermore, since $\varepsilon(\tilde{\beta}_0) = Y- \mu(\tilde{\beta}_0^{\top}X) $, it follows that $Y$ itself can also be represented by a multiple-index model. Specifically, we write $Y = m(\tilde{\gamma}, \tilde{\beta}_{l+1}^{\top}X, \dots, \tilde{\beta}_{q}^{\top}X, \varepsilon)$ with $\tilde{\gamma} \in \mathbb{R}$ and the latent projections $\tilde{\beta}_{l+1}, \dots, \tilde{\beta}_{q} \in \mathcal{S}^{p-1} $, where $q \geq l+1$. Under this framework, all these projected predictors $\tilde{\beta}_i^{\top}X$ are highly correlated with the residual $\varepsilon(\tilde{\beta}_0)$. Consequently, a natural idea is to construct the combined test statistics based on the latent projections $\tilde{\beta}_1, \dots, \tilde{\beta}_{q}$.

Note that all these latent projections are unknown and should be estimated in practice. This implies that the estimated projections are related to the test statistic itself. To address this, we randomly split the data into two parts $\mathcal{D}_1$ and $\mathcal{D}_2$ with equal sizes. Here, without loss of generality, the sample size $n$ is assumed to be even. We then use the first part of data $\mathcal{D}_1$ to estimate these projections $\tilde{\beta}_1, \dots, \tilde{\beta}_{q}$, and construct the test statistic based on the second part $\mathcal{D}_2$.
Let $\hat{\beta}_1^{(1)}, \dots, \hat{\beta}^{(1)}_{\hat{q}^{(1)}}$ be the estimators of $\beta_1, \dots, \beta_{q}$ based on $\mathcal{D}_1$, respectively.
The resulting combined test statistics are given by
\begin{align}\label{min-fisher-singletest}
\hat{T}_{\rm{min}}^{(2)} =
\min\{ \hat{p}_{\hat{\beta}^{(1)}_1}^{(2)}, \dots, \hat{p}^{(2)}_{\hat{\beta}^{(1)}_{\hat{q}^{(1)}}} \},
\quad {\rm and} \quad
\hat{T}_{\rm{Fisher}}^{(2)} = -2 \sum_{i=1}^{\hat{q}^{(1)}} \ln( \hat{p}_{\hat{\beta}^{(1)}_i}^{(2)} ),
\end{align}
where $\hat{p}_{\hat{\beta}^{(1)}_i}^{(2)} = 1- \Phi( \hat{T}_{n(2)}^{\hat{\beta}^{(1)}_i} )$ and $\hat{T}_{n(2)}^{\hat{\beta}^{(1)}_i}$ is calculated using the data from $\mathcal{D}_2$ and the estimated projection $\hat{\beta}^{(1)}_i$ from $\mathcal{D}_1$.

The aforementioned data-splitting strategy uses only half of the data to construct the test statistic, and thus the resulting test statistics may lose power under the alternatives.
Therefore, we also use data from the second part $\mathcal{D}_2$ to estimate projections and construct the test statistic based on $\mathcal{D}_1$. The resulting test statistics are given by
\begin{equation}\label{Cauchy-dualtest}
\hat{T}_{\rm{min}}^{(1)} = \min\{ \hat{p}_{\hat{\beta}^{(2)}_1}^{(1)}, \dots, \hat{p}^{(1)}_{\hat{\beta}^{(2)}_{\hat{q}^{(2)}}} \}
\quad {\rm and} \quad
\hat{T}_{\rm{Fisher}}^{(1)} = -2 \sum_{i=1}^{\hat{q}^{(2)}} \ln( \hat{p}_{\hat{\beta}^{(2)}_i}^{(1)} ),
\end{equation}
where $\hat{p}_{\hat{\beta}^{(2)}_i}^{(1)} = 1- \Phi( \hat{T}_{n(1)}^{\hat{\beta}^{(2)}_i} )$ and $\hat{T}_{n(1)}^{\hat{\beta}^{(2)}_i}$ is calculated using the data from $\mathcal{D}_1$ and the estimated projection $\hat{\beta}^{(2)}_i$ from $\mathcal{D}_2$.
Note that the test statistics $\hat{T}_{\rm min}^{(1)}$ and $\hat{T}_{\rm min}^{(2)}$ (or $\hat{T}_{\rm Fisher}^{(1)}$ and $\hat{T}_{\rm Fisher}^{(2)}$) may exhibit mutual correlation. We further construct the final test statistics using combination methods robust to dependent $p$-values, such as the Harmonic mean $p$-values \citep{Wilson2019} and the Cauchy combination approach \citep{liu2020}. In this paper, we focus on the Cauchy combination, which performs well in our simulation studies. Let $\hat{p}_{\rm min}^{(1)}$,  $\hat{p}_{\rm min}^{(2)}$,  $\hat{p}_{\rm Fisher}^{(1)}$, and  $\hat{p}_{\rm Fisher}^{(2)}$ denote the asymptotic $p$-values of $\hat{T}_{\rm min}^{(1)}$,  $\hat{T}_{\rm min}^{(2)}$,  $\hat{T}_{\rm Fisher}^{(1)}$, and $\hat{T}_{\rm Fisher}^{(2)}$, respectively. The resulting combined test statistics based on the Cauchy combination are given by
\begin{eqnarray}\label{cauchy_min_fisher}
\hat{T}_{\rm{min}}^{C} = \sum_{i=1}^2 w_i \tan\{(\frac{1}{2}-\hat{p}_{\rm{min}}^{(i)}) \pi \}
\quad {\rm and} \quad
\hat{T}_{\rm{Fisher}}^C = \sum_{i=1}^2 w_i \tan\{(\frac{1}{2}-\hat{p}_{\rm{Fisher}}^{(i)}) \pi \},
\end{eqnarray}
where the weights satisfy $\sum_{i=1}^2 w_i =1$. In this paper, we simply use the equal weights; i.e., $w_i=1/2$ for $i=1, 2$. \cite{liu2020} showed that the Cauchy combination of $p$-values under the null hypothesis can be approximated by the standard Cauchy distribution, even when these $p$-values are dependent. Consequently, the asymptotic critical values of $\hat{T}_{\rm{min}}^{C} $ and $\hat{T}_{\rm{Fisher}}^{C} $ can also be determined by quantiles of the standard Cauchy distribution. Furthermore, to obtain accurate estimates for the projections $\tilde{\beta}_1, \dots, \tilde{\beta}_{q}$, we employ model-free screening methods (e.g. DC-SIS, \cite{lirunze2012}) and sparse sufficient dimension reduction techniques (e.g. LassoSIR, \cite{lin2019}) to estimate these projections. More details on these methods are provided in the next section.

\section{Numerical studies}
\subsection{Simulation studies}
In this subsection, we investigate the finite-sample performance of the proposed tests $\hat{T}_{\rm{min}}^{C}$ and $\hat{T}_{\rm{Fisher}}^{C}$ when the dimension $p$ of the covariates $X$ may exceed the sample size $n$. To compute $\hat{T}_{\rm{min}}^{C}$ and $\hat{T}_{\rm{Fisher}}^{C}$, we follow the suggestion of \cite{patilea2016} and use the Epanechnikov kernel $K(x) = (3/4)(1-x^2)I(|x| \leq 1)$ and a bandwidth $h = c_h n^{-2/9}$ with $c_h \in \{ 0.75, 1.00, 1.25\}$. The choice of different bandwidths allows us to investigate their impact on the empirical size and power of the proposed tests. To obtain an accurate estimate of $\beta_0$ under the null, we consider the post-Lasso estimator of $\beta_0$, which applies the least squares to the model selected by the Lasso estimator. As shown by \cite{Chernozhukov2013}, the least squares post-Lasso performs at least as well as Lasso in terms of the rate of convergence, and has the advantage of a smaller bias. Consequently, the post-Lasso estimator $\hat{\beta}_0$ satisfies Condition (A1) imposed in Section 3 of this paper. We also conducted simulations for our proposed tests based on Lasso estimators. These unreported results show that the tests fail to maintain the significance level unless the covariate dimension $p$ is small relative to the sample size $n$. This is likely attributable to the inherent bias of the Lasso estimator in high dimensional settings, which causes a substantial deviation of the estimator from the true parameter $\beta_0$.

We compare our tests with the recent high dimensional goodness-of-fit tests, $RP_n$ and $GRP_n$, proposed by \cite{Shah2018} and \cite{Jankova2020}, respectively. The simulation results are based on the average of 1000 replications with a significance level of $\tau=0.05$. The parameter $a=0$ corresponds to the null hypothesis, while $a \neq 0$ corresponds to the alternative hypotheses. The simulation results of the $RP_n$ test are obtained using the R package RPtests, while those for $GRP_n$ are obtained by running the code available on the website \url{https://github.com/jankova/GRPtests}, which was posted by the authors of \cite{Jankova2020}.

In the first simulation study, we consider the case of testing Gaussian linear models in high dimensional settings.

{\em Study 1.} Generate data from the following models:
\begin{eqnarray*}
H_{11}:  Y &=& \beta_0^{\top}X+ 0.1a(\beta_0^{\top}X)^2 + \varepsilon;  \\
H_{12}:  Y &=& \beta_0^{\top}X+ a \cos(0.6 \pi \beta_0^{\top}X) + \varepsilon;\\
H_{13}:  Y &=& \beta_0^{\top}X+ a \exp(0.5 \beta_1^{\top}X) + \varepsilon;
\end{eqnarray*}
where $\beta_0 = (1,1,1,1,1,0,\dots,0)^{\top}$ and $\beta_1=(\underbrace{1,\dots,1}_{p_1},0,\dots,0)$ with $p_1 = 10$. The predictor vector $X$ is $N(0, \Sigma)$ independent of the standard Gaussian error term $\varepsilon$, where $\Sigma = I_p$ or $\Sigma=(\rho^{|i-j|})_{p \times p}$ with $\rho = 0.4$ or $\rho = 0.8$. We consider the sample size $n=300$ and the dimension $p \in \{50, 100, 300, 600, 900, 1200\}$. Note that under the alternatives, $H_{12}$ is a high-frequency model, while $H_{11}$ and $H_{13}$ are low-frequency models. To compute our tests $\hat{T}_{\rm{min}}^{C}$ and $\hat{T}_{\rm{Fisher}}^{C}$, we randomly split the data into two parts, $\mathcal{D}_1$ and $\mathcal{D}_2$, and then construct the test statistics according to (\ref{cauchy_min_fisher}). The sparse sufficient dimension reduction method LassoSIR \citep{lin2019} may not perform well when the covariate dimension $p$ is too large. Therefore, we use the Distance Correlation Sure Independence Screening \citep[DC-SIS]{lirunze2012} to select the first $\frac{n/2}{\log(n/2)}$ top-ranked variables. We then apply LassoSIR to these selected variables to construct the estimated projections $\hat{\beta}_i^{(1)}$ and $\hat{\beta}_i^{(2)}$. Furthermore, despite the null models having zero intercepts, we treat them as unknown parameters in the simulation studies.

Table 1-3 present the empirical sizes and powers of our tests $\hat{T}_{\rm{min}}^{C}$ and $\hat{T}_{\rm{Fisher}}^{C}$ for different bandwidths $h = c_h n^{-2/9}$ with $c_h \in \{ 0.75, 1.00, 1.25 \}$ under the null and alternative hypotheses. We observe that our tests maintain the significance level very well for all dimensions and sample sizes. Under the alternatives, our tests exhibit good power in most cases. Moreover, these results showcase that our tests are less affected by the bandwidth, even when the dimension of the covariates $X$ is much larger than the sample size.

Tables 4-6 present the simulation results of $RP_n$, $GRP_n$, and our tests $\hat{T}_{\rm{min}}^{C}$ and $\hat{T}_{\rm{Fisher}}^{C}$ using the bandwidth $h = n^{-2/9}$. It can be observed that $RP_n$ and our tests $\hat{T}_{Cauchy}$ and $\hat{T}_{Cauchy}^D$ can maintain the significance level very well in all cases, whereas the $GRP_n$ test is slightly conservative with smaller empirical sizes when the dimension $p$ is large. For the empirical power, we can see that the proposed tests $\hat{T}_{\rm{min}}^{C}$ and $\hat{T}_{\rm{Fisher}}^{C}$ typically have higher power than $RP_n$ and $GRP_n$, particularly for the high frequency model $H_{12}$. Note that both $RP_n$ and $GRP_n$ have almost no power for model $H_{12}$ across all dimension and correlation settings. Interestingly, the empirical powers of all these tests appear to increase as the correlation of the covariates $X$ increases.

\begin{table}[ht!]\caption{Empirical sizes and powers of our tests $\hat{T}_{\rm{min}}^C$ and $\hat{T}_{\rm{Fisher}}^C$ with different bandwidths for $H_{11}$ in Study 1.}\label{table-bandwidth-H11}
\centering
{\small\scriptsize\hspace{8cm}
\renewcommand{\arraystretch}{0.6}\tabcolsep 0.4cm
\begin{tabular}{*{20}{c}}
\hline
&\multicolumn{1}{c}{$c_h$}&\multicolumn{1}{c}{n=300}&\multicolumn{1}{c}{n=300}&\multicolumn{1}{c}{n=300}&\multicolumn{1}{c}{n=300}
&\multicolumn{1}{c}{n=300}&\multicolumn{1}{c}{n=300}
\\
&&\multicolumn{1}{c}{p=50}&\multicolumn{1}{c}{p=100}&\multicolumn{1}{c}{p=300}&\multicolumn{1}{c}{p=600}
&\multicolumn{1}{c}{p=900}&\multicolumn{1}{c}{p=1200} \\
\hline
$\hat{T}_{\rm{min}}^C, \ \Sigma = I_p, \ a = 0$
& 0.75 &0.051 &0.054 &0.049 &0.050 &0.054 &0.061\\
& 1.00 &0.032 &0.033 &0.059 &0.057 &0.045 &0.058\\
& 1.25 &0.042 &0.055 &0.045 &0.036 &0.043 &0.054\\
\\
$\hat{T}^C_{\rm{Fisher}}, \ \Sigma = I_p, \ a = 0$
& 0.75 &0.028 &0.042 &0.029 &0.039 &0.048 &0.046\\
& 1.00 &0.019 &0.025 &0.041 &0.045 &0.033 &0.041\\
& 1.25 &0.013 &0.035 &0.023 &0.022 &0.031 &0.034\\
\\
$\hat{T}_{\rm{min}}^C, \ \Sigma = I_p, \ a = 1$
& 0.75 &0.919 &0.819 &0.581 &0.439 &0.409 &0.344\\
& 1.00 &0.957 &0.872 &0.641 &0.529 &0.438 &0.394\\
& 1.25 &0.974 &0.905 &0.684 &0.563 &0.482 &0.435\\
\\
$\hat{T}^C_{\rm{Fisher}}, \ \Sigma = I_p, \ a = 1$
& 0.75 &0.870 &0.725 &0.501 &0.386 &0.347 &0.295\\
& 1.00 &0.909 &0.793 &0.562 &0.437 &0.388 &0.336\\
& 1.25 &0.945 &0.856 &0.608 &0.509 &0.436 &0.379\\
\hline
$\hat{T}_{\rm{min}}^C, \ \Sigma = (0.4^{|i-j|})_{p\times p}, \ a = 0$
& 0.75 &0.048 &0.068 &0.056 &0.046 &0.066 &0.054\\
& 1.00 &0.046 &0.045 &0.059 &0.057 &0.046 &0.048\\
& 1.25 &0.047 &0.049 &0.058 &0.045 &0.044 &0.048\\
\\
$\hat{T}^C_{\rm{Fisher}}, \ \Sigma = (0.4^{|i-j|})_{p\times p}, \ a = 0$
& 0.75 &0.020 &0.043 &0.028 &0.030 &0.044 &0.043 \\
& 1.00 &0.031 &0.025 &0.039 &0.037 &0.030 &0.030 \\
& 1.25 &0.022 &0.029 &0.025 &0.029 &0.029 &0.026\\
\\
$\hat{T}_{\rm{min}}^C, \ \Sigma = (0.4^{|i-j|})_{p\times p}, \ a = 1$
& 0.75 &1.000 &1.000 &0.981 &0.944 &0.899 &0.877\\
& 1.00 &1.000 &1.000 &0.987 &0.940 &0.910 &0.899\\
& 1.25 &1.000 &1.000 &0.994 &0.976 &0.935 &0.901\\
\\
$\hat{T}^C_{\rm{Fisher}}, \ \Sigma = (0.4^{|i-j|})_{p\times p}, \ a = 1$
& 0.75 &1.000 &0.998 &0.971 &0.917 &0.868 &0.845\\
& 1.00 &1.000 &1.000 &0.976 &0.927 &0.897 &0.885\\
& 1.25 &1.000 &1.000 &0.988 &0.964 &0.909 &0.878\\
\hline
$\hat{T}_{\rm{min}}^C, \ \Sigma = (0.8^{|i-j|})_{p\times p}, \ a = 0$
& 0.75 &0.052 &0.056 &0.064 &0.047 &0.053 &0.056\\
& 1.00 &0.049 &0.038 &0.054 &0.058 &0.059 &0.050\\
& 1.25 &0.037 &0.047 &0.046 &0.038 &0.049 &0.050\\
\\
$\hat{T}^C_{\rm{Fisher}}, \ \Sigma = (0.8^{|i-j|})_{p\times p}, \ a = 0$
& 0.75 &0.037 &0.030 &0.037 &0.029 &0.041 &0.047 \\
& 0.75 &0.028 &0.022 &0.031 &0.031 &0.037 &0.038 \\
& 0.75 &0.023 &0.025 &0.026 &0.019 &0.023 &0.039 \\
\\
$\hat{T}_{\rm{min}}^C, \ \Sigma = (0.8^{|i-j|})_{p\times p}, \ a = 1$
& 0.75 &1.000 &1.000 &1.000 &0.999 &0.995 &0.986\\
& 1.00 &1.000 &1.000 &1.000 &0.999 &0.995 &0.992\\
& 1.25 &1.000 &1.000 &1.000 &0.998 &0.996 &0.996\\
\\
$\hat{T}^C_{\rm{Fisher}}, \ \Sigma = (0.8^{|i-j|})_{p\times p}, \ a = 1$
& 0.75 &1.000 &1.000 &1.000 &0.996 &0.992 &0.981\\
& 1.00 &1.000 &1.000 &1.000 &0.999 &0.994 &0.989\\
& 1.25 &1.000 &1.000 &1.000 &0.998 &0.994 &0.995\\
\hline
\end{tabular}}
\end{table}

\begin{table}[ht!]\caption{Empirical sizes and powers of our tests $\hat{T}_{\rm{min}}^C$ and $\hat{T}^C_{\rm{Fisher}}$ with different bandwidths for $H_{12}$ in Study 1.}\label{table-bandwidth-H12}
\centering
{\small\scriptsize\hspace{8cm}
\renewcommand{\arraystretch}{0.6}\tabcolsep 0.4cm
\begin{tabular}{*{20}{c}}
\hline
&\multicolumn{1}{c}{$c_h$}&\multicolumn{1}{c}{n=300}&\multicolumn{1}{c}{n=300}&\multicolumn{1}{c}{n=300}&\multicolumn{1}{c}{n=300}
&\multicolumn{1}{c}{n=300}&\multicolumn{1}{c}{n=300}
\\
&&\multicolumn{1}{c}{p=50}&\multicolumn{1}{c}{p=100}&\multicolumn{1}{c}{p=300}&\multicolumn{1}{c}{p=600}
&\multicolumn{1}{c}{p=900}&\multicolumn{1}{c}{p=1200} \\
\hline
$\hat{T}_{\rm{min}}^C, \ \Sigma = I_p, \ a = 0$
& 0.75 &0.047 &0.054 &0.061 &0.059 &0.059 &0.058\\
& 1.00 &0.043 &0.047 &0.057 &0.054 &0.053 &0.063\\
& 1.25 &0.036 &0.050 &0.049 &0.043 &0.043 &0.049\\
\\
$\hat{T}^C_{\rm{Fisher}}, \ \Sigma = I_p, \ a = 0$
& 0.75 &0.029 &0.032 &0.041 &0.044 &0.043 &0.051\\
& 1.00 &0.022 &0.021 &0.033 &0.035 &0.041 &0.045\\
& 1.25 &0.018 &0.026 &0.028 &0.030 &0.029 &0.028\\
\\
$\hat{T}_{\rm{min}}^C, \ \Sigma = I_p, \ a = 1$
& 0.75 &0.877 &0.706 &0.378 &0.217 &0.162 &0.150\\
& 1.00 &0.882 &0.725 &0.425 &0.239 &0.181 &0.167\\
& 1.25 &0.883 &0.722 &0.385 &0.230 &0.145 &0.130\\
\\
$\hat{T}^C_{\rm{Fisher}}, \ \Sigma = I_p, \ a = 1$
& 0.75 &0.834 &0.634 &0.306 &0.176 &0.130 &0.139\\
& 1.00 &0.853 &0.661 &0.355 &0.201 &0.141 &0.127\\
& 1.25 &0.847 &0.635 &0.320 &0.194 &0.113 &0.107\\
\hline
$\hat{T}_{\rm{min}}^C, \ \Sigma = (0.4^{|i-j|})_{p\times p}, \ a = 0$
& 0.75 &0.047 &0.048 &0.060 &0.071 &0.051 &0.049\\
& 1.00 &0.041 &0.043 &0.068 &0.050 &0.056 &0.068\\
& 1.25 &0.026 &0.043 &0.054 &0.036 &0.053 &0.041\\
\\
$\hat{T}^C_{\rm{Fisher}}, \ \Sigma = (0.4^{|i-j|})_{p\times p}, \ a = 0$
& 0.75 &0.027 &0.040 &0.028 &0.050 &0.037 &0.034\\
& 1.00 &0.020 &0.030 &0.046 &0.036 &0.037 &0.045\\
& 1.25 &0.015 &0.028 &0.027 &0.023 &0.033 &0.031\\
\\
$\hat{T}_{\rm{min}}^C, \ \Sigma = (0.4^{|i-j|})_{p\times p}, \ a = 1$
& 0.75 &0.945 &0.861 &0.612 &0.448 &0.391 &0.308\\
& 1.00 &0.933 &0.875 &0.626 &0.480 &0.390 &0.338\\
& 1.25 &0.943 &0.842 &0.618 &0.471 &0.374 &0.357\\
\\
$\hat{T}^C_{\rm{Fisher}}, \ \Sigma = (0.4^{|i-j|})_{p\times p}, \ a = 1$
& 0.75 &0.920 &0.798 &0.546 &0.397 &0.334 &0.278\\
& 1.00 &0.909 &0.823 &0.551 &0.427 &0.334 &0.296\\
& 1.25 &0.909 &0.781 &0.551 &0.405 &0.320 &0.311\\
\hline
$\hat{T}_{\rm{min}}^C, \ \Sigma = (0.8^{|i-j|})_{p\times p}, \ a = 0$
& 0.75 &0.041 &0.049 &0.062 &0.046 &0.057 &0.051\\
& 1.00 &0.051 &0.047 &0.049 &0.042 &0.061 &0.046\\
& 1.25 &0.036 &0.040 &0.052 &0.050 &0.052 &0.055\\
\\
$\hat{T}^C_{\rm{Fisher}}, \ \Sigma = (0.8^{|i-j|})_{p\times p}, \ a = 0$
& 0.75 &0.028 &0.027 &0.039 &0.027 &0.032 &0.035\\
& 1.00 &0.031 &0.029 &0.026 &0.030 &0.038 &0.028\\
& 1.25 &0.019 &0.025 &0.027 &0.027 &0.030 &0.040\\
\\
$\hat{T}_{\rm{min}}^C, \ \Sigma = (0.8^{|i-j|})_{p\times p}, \ a = 1$
& 0.75 &0.919 &0.876 &0.795 &0.720 &0.659 &0.629\\
& 1.00 &0.906 &0.886 &0.810 &0.741 &0.685 &0.622\\
& 1.25 &0.908 &0.891 &0.782 &0.719 &0.645 &0.644\\
\\
$\hat{T}^C_{\rm{Fisher}}, \ \Sigma = (0.8^{|i-j|})_{p\times p}, \ a = 1$
& 0.75 &0.876 &0.831 &0.731 &0.651 &0.597 &0.565\\
& 1.00 &0.879 &0.843 &0.753 &0.696 &0.612 &0.578\\
& 1.25 &0.874 &0.853 &0.729 &0.639 &0.571 &0.580\\
\hline
\end{tabular}}
\end{table}

\begin{table}[ht!]\caption{Empirical sizes and powers of our tests $\hat{T}_{\rm{min}}^C$ and $\hat{T}^C_{\rm{Fisher}}$ with different bandwidths for $H_{13}$ in Study 1.}\label{table-bandwidth-H13}
\centering
{\small\scriptsize\hspace{8cm}
\renewcommand{\arraystretch}{0.6}\tabcolsep 0.4cm
\begin{tabular}{*{20}{c}}
\hline
&\multicolumn{1}{c}{$c_h$}&\multicolumn{1}{c}{n=300}&\multicolumn{1}{c}{n=300}&\multicolumn{1}{c}{n=300}&\multicolumn{1}{c}{n=300}
&\multicolumn{1}{c}{n=300}&\multicolumn{1}{c}{n=300}
\\
&&\multicolumn{1}{c}{p=50}&\multicolumn{1}{c}{p=100}&\multicolumn{1}{c}{p=300}&\multicolumn{1}{c}{p=600}
&\multicolumn{1}{c}{p=900}&\multicolumn{1}{c}{p=1200} \\
\hline
$\hat{T}_{\rm{min}}^C, \ \Sigma = I_p, \ a = 0$
& 0.75 &0.047 &0.053 &0.048 &0.066 &0.051 &0.053\\
& 1.00 &0.049 &0.056 &0.061 &0.054 &0.056 &0.062\\
& 1.25 &0.039 &0.040 &0.058 &0.063 &0.055 &0.044\\
\\
$\hat{T}^C_{\rm{Fisher}}, \ \Sigma = I_p, \ a = 0$
& 0.75 &0.027 &0.030 &0.037 &0.043 &0.036 &0.042\\
& 1.00 &0.030 &0.028 &0.032 &0.036 &0.030 &0.037\\
& 1.25 &0.024 &0.022 &0.032 &0.039 &0.039 &0.039\\
\\
$\hat{T}_{\rm{min}}^C, \ \Sigma = I_p, \ a = 1$
& 0.75 &0.993 &0.920 &0.570 &0.394 &0.328 &0.318\\
& 1.00 &0.995 &0.937 &0.636 &0.447 &0.389 &0.360\\
& 1.25 &0.998 &0.956 &0.656 &0.476 &0.409 &0.390\\
\\
$\hat{T}^C_{\rm{Fisher}}, \ \Sigma = I_p, \ a = 1$
& 0.75 &0.987 &0.881 &0.489 &0.348 &0.308 &0.275\\
& 1.00 &0.993 &0.892 &0.562 &0.395 &0.365 &0.310\\
& 1.25 &0.996 &0.931 &0.564 &0.413 &0.344 &0.339\\
\hline
$\hat{T}_{\rm{min}}^C, \ \Sigma = (0.4^{|i-j|})_{p\times p}, \ a = 0$
& 0.75 &0.054 &0.055 &0.052 &0.052 &0.059 &0.052\\
& 1.00 &0.054 &0.037 &0.062 &0.053 &0.044 &0.052\\
& 1.25 &0.040 &0.046 &0.043 &0.049 &0.047 &0.049\\
\\
$\hat{T}^C_{\rm{Fisher}}, \ \Sigma = (0.4^{|i-j|})_{p\times p}, \ a = 0$
& 0.75 &0.034 &0.033 &0.028 &0.031 &0.043 &0.041\\
& 1.00 &0.032 &0.022 &0.041 &0.034 &0.036 &0.041\\
& 1.25 &0.025 &0.024 &0.020 &0.034 &0.031 &0.033\\
\\
$\hat{T}_{\rm{min}}^C, \ \Sigma = (0.4^{|i-j|})_{p\times p}, \ a = 1$
& 0.75 &0.998 &0.989 &0.943 &0.858 &0.796 &0.769\\
& 1.00 &1.000 &0.995 &0.952 &0.900 &0.859 &0.803\\
& 1.25 &0.999 &0.997 &0.979 &0.935 &0.898 &0.850\\
\\
$\hat{T}^C_{\rm{Fisher}}, \ \Sigma = (0.4^{|i-j|})_{p\times p}, \ a = 1$
& 0.75 &0.998 &0.982 &0.902 &0.807 &0.756 &0.719\\
& 1.00 &1.000 &0.987 &0.921 &0.860 &0.822 &0.782\\
& 1.25 &0.999 &0.994 &0.953 &0.892 &0.861 &0.806\\
\hline
\\
$\hat{T}_{\rm{min}}^C, \ \Sigma = (0.8^{|i-j|})_{p\times p}, \ a = 0$
& 0.75 &0.038 &0.052 &0.052 &0.061 &0.067 &0.063\\
& 1.00 &0.048 &0.041 &0.049 &0.061 &0.050 &0.047\\
& 1.25 &0.035 &0.030 &0.038 &0.054 &0.053 &0.046\\
\\
$\hat{T}^C_{\rm{Fisher}}, \ \Sigma = (0.8^{|i-j|})_{p\times p}, \ a = 0$
& 0.75 &0.025 &0.028 &0.030 &0.043 &0.049 &0.037\\
& 1.00 &0.025 &0.031 &0.031 &0.042 &0.034 &0.032\\
& 1.25 &0.020 &0.017 &0.020 &0.033 &0.035 &0.028\\
\\
$\hat{T}_{\rm{min}}^C, \ \Sigma = (0.8^{|i-j|})_{p\times p}, \ a = 1$
& 0.75 &0.999 &0.998 &1.000 &0.989 &0.980 &0.979\\
& 1.00 &1.000 &0.998 &1.000 &0.992 &0.985 &0.982\\
& 1.25 &1.000 &1.000 &0.999 &0.997 &0.983 &0.990\\
\\
$\hat{T}^C_{\rm{Fisher}}, \ \Sigma = (0.8^{|i-j|})_{p\times p}, \ a = 1$
& 0.75 &1.000 &0.999 &1.000 &0.982 &0.974 &0.979\\
& 1.00 &1.000 &0.999 &0.999 &0.987 &0.988 &0.977\\
& 1.25 &1.000 &1.000 &0.999 &0.994 &0.978 &0.991\\
\hline
\end{tabular}}
\end{table}

\begin{table}[ht!]\caption{Empirical sizes and powers of the tests $RP_n$, $GRP_n$, $\hat{T}_{\rm{min}}^C$, and $\hat{T}_{\rm{Fisher}}^C$ with bandwidth $h = n^{-2/9}$ for $H_{11}$ in Study 1.}\label{table-H11}
\centering
{\small\scriptsize\hspace{8cm}
\renewcommand{\arraystretch}{0.6}\tabcolsep 0.4cm
\begin{tabular}{*{20}{c}}
\hline
&\multicolumn{1}{c}{a}&\multicolumn{1}{c}{n=300}&\multicolumn{1}{c}{n=300}&\multicolumn{1}{c}{n=300}&\multicolumn{1}{c}{n=300}
&\multicolumn{1}{c}{n=300}&\multicolumn{1}{c}{n=300}\\
&&\multicolumn{1}{c}{p=50}&\multicolumn{1}{c}{p=100}&\multicolumn{1}{c}{p=300}&\multicolumn{1}{c}{p=600}
&\multicolumn{1}{c}{p=900}&\multicolumn{1}{c}{p=1200} \\
\hline
$\hat{T}_{\rm{min}}^C, \ \Sigma = I_p$
& 0.0 &0.046 &0.047 &0.053 &0.059 &0.046 &0.043\\
& 1.0 &0.955 &0.884 &0.611 &0.500 &0.467 &0.420\\
\\
$\hat{T}^C_{\rm{Fisher}}, \ \Sigma = I_p$
& 0.0 &0.024 &0.022 &0.027 &0.050 &0.030 &0.035\\
& 1.0 &0.914 &0.821 &0.532 &0.437 &0.425 &0.373\\
\\
$RP_n, \ \Sigma = I_p$
& 0.0 &0.032 &0.045 &0.040 &0.039 &0.032 &0.037\\
& 1.0 &0.161 &0.165 &0.118 &0.134 &0.104 &0.097\\
\\
$GRP_n, \ \Sigma = I_p$
& 0.0 &0.030 &0.034 &0.016 &0.010 &0.014 &0.011\\
& 1.0 &0.264 &0.183 &0.072 &0.058 &0.045 &0.045\\
\hline
$\hat{T}_{\rm{min}}^C, \ \Sigma = (0.4^{|i-j|})_{p\times p}$
& 0.0 &0.042 &0.057 &0.063 &0.049 &0.044 &0.048\\
& 1.0 &1.000 &1.000 &0.988 &0.961 &0.934 &0.902\\
\\
$\hat{T}^C_{\rm{Fisher}}, \ \Sigma = (0.4^{|i-j|})_{p\times p}$
& 0.0 &0.020 &0.035 &0.039 &0.031 &0.031 &0.030\\
& 1.0 &1.000 &1.000 &0.983 &0.940 &0.912 &0.890\\
\\
$RP_n, \ \Sigma = (0.4^{|i-j|})_{p\times p}$
& 0.0 &0.037 &0.034 &0.049 &0.035 &0.046 &0.034\\
& 1.0 &0.544 &0.523 &0.478 &0.455 &0.458 &0.422\\
\\
$GRP_n, \ \Sigma = (0.4^{|i-j|})_{p\times p}$
& 0.0 &0.036 &0.029 &0.026 &0.015 &0.013 &0.014\\
& 1.0 &0.998 &0.977 &0.703 &0.574 &0.541 &0.482\\
\hline
$\hat{T}_{\rm{min}}^C, \ \Sigma = (0.8^{|i-j|})_{p\times p}$
& 0.0 &0.036 &0.053 &0.057 &0.044 &0.045 &0.055\\
& 1.0 &1.000 &1.000 &0.999 &0.999 &0.998 &0.992\\
\\
$\hat{T}^C_{\rm{Fisher}}, \ \Sigma = (0.8^{|i-j|})_{p\times p}$
& 0.0 &0.022 &0.031 &0.031 &0.028 &0.029 &0.040\\
& 1.0 &1.000 &1.000 &0.999 &0.998 &0.996 &0.990\\
\\
$RP_n, \ \Sigma = (0.8^{|i-j|})_{p\times p}$
& 0.0 &0.043 &0.055 &0.041 &0.042 &0.029 &0.031\\
& 1.0 &0.745 &0.677 &0.605 &0.568 &0.555 &0.555\\
\\
$GRP_n, \ \Sigma = (0.8^{|i-j|})_{p\times p}$
& 0.0 &0.044 &0.035 &0.033 &0.024 &0.031 &0.019\\
& 1.0 &1.000 &1.000 &1.000 &0.993 &0.996 &0.995\\
\hline
\end{tabular}}
\end{table}

\begin{table}[ht!]\caption{Empirical sizes and powers of the tests $RP_n$, $GRP_n$, $\hat{T}_{\rm{min}}^C$, and $\hat{T}_{\rm{Fisher}}^C$ with bandwidth $h = n^{-2/9}$ for $H_{12}$ in Study 1.}\label{table-H12}
\centering
{\small\scriptsize\hspace{8cm}
\renewcommand{\arraystretch}{0.6}\tabcolsep 0.4cm
\begin{tabular}{*{20}{c}}
\hline
&\multicolumn{1}{c}{a}&\multicolumn{1}{c}{n=300}&\multicolumn{1}{c}{n=300}&\multicolumn{1}{c}{n=300}&\multicolumn{1}{c}{n=300}
&\multicolumn{1}{c}{n=300}&\multicolumn{1}{c}{n=300}\\
&&\multicolumn{1}{c}{p=50}&\multicolumn{1}{c}{p=100}&\multicolumn{1}{c}{p=300}&\multicolumn{1}{c}{p=600}
&\multicolumn{1}{c}{p=900}&\multicolumn{1}{c}{p=1200} \\
\hline
$\hat{T}_{\rm{min}}^C, \ \Sigma = I_p$
& 0.0 &0.038 &0.055 &0.063 &0.043 &0.051 &0.047\\
& 1.0 &0.883 &0.735 &0.392 &0.255 &0.197 &0.160\\
\\
$\hat{T}^C_{\rm{Fisher}}, \ \Sigma = I_p$
& 0.0 &0.022 &0.028 &0.024 &0.031 &0.030 &0.040\\
& 1.0 &0.844 &0.680 &0.320 &0.202 &0.161 &0.121\\
\\
$RP_n, \ \Sigma = I_p$
& 0.0 &0.040 &0.040 &0.043 &0.034 &0.026 &0.041\\
& 1.0 &0.041 &0.036 &0.034 &0.036 &0.037 &0.037\\
\\
$GRP_n, \ \Sigma = I_p$
& 0.0 &0.029 &0.027 &0.021 &0.016 &0.008 &0.010\\
& 1.0 &0.074 &0.063 &0.043 &0.033 &0.024 &0.018\\
\hline
$\hat{T}_{\rm{min}}^C, \ \Sigma = (0.4^{|i-j|})_{p\times p}$
& 0.0 &0.046 &0.042 &0.068 &0.043 &0.053 &0.056\\
& 1.0 &0.940 &0.866 &0.631 &0.513 &0.382 &0.325\\
\\
$\hat{T}^C_{\rm{Fisher}}, \ \Sigma = (0.4^{|i-j|})_{p\times p}$
& 0.0 &0.029 &0.025 &0.043 &0.028 &0.039 &0.038\\
& 1.0 &0.906 &0.814 &0.548 &0.451 &0.330 &0.275\\
\\
$RP_n, \ \Sigma = (0.4^{|i-j|})_{p\times p}$
& 0.0 &0.034 &0.040 &0.028 &0.033 &0.041 &0.043\\
& 1.0 &0.044 &0.049 &0.034 &0.038 &0.034 &0.036\\
\\
$GRP_n, \ \Sigma = (0.4^{|i-j|})_{p\times p}$
& 0.0 &0.029 &0.036 &0.021 &0.023 &0.016 &0.017\\
& 1.0 &0.078 &0.067 &0.056 &0.038 &0.047 &0.034\\
\hline
$\hat{T}_{\rm{min}}^C, \ \Sigma = (0.8^{|i-j|})_{p\times p}$
& 0.0 &0.042 &0.052 &0.048 &0.043 &0.057 &0.043\\
& 1.0 &0.921 &0.882 &0.804 &0.729 &0.688 &0.640\\
\\
$\hat{T}^C_{\rm{Fisher}}, \ \Sigma = (0.8^{|i-j|})_{p\times p}$
& 0.0 &0.024 &0.021 &0.026 &0.031 &0.049 &0.030\\
& 1.0 &0.884 &0.846 &0.738 &0.672 &0.614 &0.573\\
\\
$RP_n, \ \Sigma = (0.8^{|i-j|})_{p\times p}$
& 0.0 &0.036 &0.044 &0.053 &0.040 &0.036 &0.033\\
& 1.0 &0.030 &0.040 &0.035 &0.038 &0.032 &0.029\\
\\
$GRP_n, \ \Sigma = (0.8^{|i-j|})_{p\times p}$
& 0.0 &0.029 &0.032 &0.030 &0.033 &0.024 &0.032\\
& 1.0 &0.084 &0.066 &0.055 &0.067 &0.054 &0.070\\
\hline
\end{tabular}}
\end{table}

\begin{table}[ht!]\caption{Empirical sizes and powers of the tests $RP_n$, $GRP_n$, $\hat{T}_{\rm{min}}^C$, and $\hat{T}_{\rm{Fisher}}^C$ with bandwidth $h = n^{-2/9}$ for $H_{13}$ in Study 1.}\label{table-H13}
\centering
{\small\scriptsize\hspace{8cm}
\renewcommand{\arraystretch}{0.6}\tabcolsep 0.4cm
\begin{tabular}{*{20}{c}}
\hline
&\multicolumn{1}{c}{a}&\multicolumn{1}{c}{n=300}&\multicolumn{1}{c}{n=300}&\multicolumn{1}{c}{n=300}&\multicolumn{1}{c}{n=300}
&\multicolumn{1}{c}{n=300}&\multicolumn{1}{c}{n=300}\\
&&\multicolumn{1}{c}{p=50}&\multicolumn{1}{c}{p=100}&\multicolumn{1}{c}{p=300}&\multicolumn{1}{c}{p=600}
&\multicolumn{1}{c}{p=900}&\multicolumn{1}{c}{p=1200} \\
\hline
$\hat{T}_{\rm{min}}^C, \ \Sigma = I_p$
& 0.0 &0.045 &0.051 &0.043 &0.049 &0.055 &0.054\\
& 1.0 &0.993 &0.955 &0.612 &0.460 &0.386 &0.370\\
\\
$\hat{T}^C_{\rm{Fisher}}, \ \Sigma = I_p$
& 0.0 &0.024 &0.030 &0.022 &0.034 &0.036 &0.040\\
& 1.0 &0.987 &0.919 &0.553 &0.406 &0.345 &0.319\\
\\
$RP_n, \ \Sigma = I_p$
& 0.0 &0.038 &0.050 &0.030 &0.031 &0.032 &0.029\\
& 1.0 &0.106 &0.093 &0.106 &0.095 &0.099 &0.103\\
\\
$GRP_n, \ \Sigma = I_p$
& 0.0 &0.034 &0.029 &0.028 &0.022 &0.018 &0.013\\
& 1.0 &0.698 &0.578 &0.395 &0.389 &0.395 &0.384\\
\hline
$\hat{T}_{\rm{min}}^C, \ \Sigma = (0.4^{|i-j|})_{p\times p}$
& 0.0 &0.058 &0.057 &0.052 &0.055 &0.041 &0.050\\
& 1.0 &0.999 &0.997 &0.959 &0.898 &0.863 &0.826\\
\\
$\hat{T}^C_{\rm{Fisher}}, \ \Sigma = (0.4^{|i-j|})_{p\times p}$
& 0.0 &0.040 &0.029 &0.031 &0.028 &0.026 &0.040\\
& 1.0 &0.998 &0.995 &0.929 &0.864 &0.820 &0.789\\
\\
$RP_n, \ \Sigma = (0.4^{|i-j|})_{p\times p}$
& 0.0 &0.049 &0.037 &0.043 &0.041 &0.032 &0.036\\
& 1.0 &0.184 &0.158 &0.167 &0.179 &0.140 &0.186\\
\\
$GRP_n, \ \Sigma = (0.4^{|i-j|})_{p\times p}$
& 0.0 &0.041 &0.035 &0.018 &0.022 &0.023 &0.025\\
& 1.0 &0.975 &0.928 &0.819 &0.768 &0.730 &0.689\\
\hline
$\hat{T}_{\rm{min}}^C, \ \Sigma = (0.8^{|i-j|})_{p\times p}$
& 0.0 &0.048 &0.053 &0.053 &0.073 &0.043 &0.057\\
& 1.0 &1.000 &0.999 &0.999 &0.996 &0.985 &0.984\\
\\
$\hat{T}^C_{\rm{Fisher}}, \ \Sigma = (0.8^{|i-j|})_{p\times p}$
& 0.0 &0.033 &0.033 &0.031 &0.042 &0.031 &0.042\\
& 1.0 &1.000 &0.999 &0.997 &0.993 &0.991 &0.977\\
\\
$RP_n, \ \Sigma = (0.8^{|i-j|})_{p\times p}$
& 0.0 &0.037 &0.042 &0.039 &0.042 &0.050 &0.045\\
& 1.0 &0.391 &0.348 &0.311 &0.330 &0.320 &0.311\\
\\
$GRP_n, \ \Sigma = (0.8^{|i-j|})_{p\times p}$
& 0.0 &0.050 &0.031 &0.028 &0.026 &0.031 &0.034\\
& 1.0 &1.000 &0.999 &0.998 &0.988 &0.978 &0.983\\
\hline
\end{tabular}}
\end{table}

We have considered goodness-of-fit tests for Gaussian linear models in Study 1. Next, we assess the performance of the proposed tests for the goodness-of-fit for generalized linear models in high dimensional scenarios.

{\em Study 2.} The data are generated from the logistic regression model according to
$$ Y|X \sim Bernoulli\{ \mu(\beta_0^{\top}X + ag(X)) \},$$
where $\mu(z) = 1/(1+\exp(-z))$. We consider two different cases for the misspecified $g(X)$:
\begin{eqnarray*}
H_{21}:  g(X) &=& 0.2(\beta_0^{\top}X)^2,   \\
H_{22}:  g(X) &=& X^{(1)} X^{(2)} + X^{(2)}X^{(3)} + X^{(3)}X^{(4)} + X^{(4)}X^{(5)},
\end{eqnarray*}
where the parameter $\beta_0 = (1,1,1,1,1,0,\dots,0)^{\top}$ and the predictor vector $X$ are the same with study 1, and the sample size $n=600$ with dimension $p \in \{50, 100, 300, 600, 900, 1200\}$.

The empirical sizes and powers of our tests $\hat{T}_{\rm{min}}^{C}$ and $\hat{T}_{\rm{Fisher}}^{C}$ with different bandwidths are presented in Table 7-8. These results indicate that our tests can control the nominal significance level in most cases. However, in the case of high correlation of the predictor vector $X$, the test $\hat{T}_{\rm{min}}^{C}$ becomes liberal with large empirical sizes when the covariate dimension $p$ is larger than the sample size $n$. Consistent with findings for testing linear models, the empirical powers also increase as the correlation of the predictor vector grows.

Since the $RP_n$ test cannot be applied to test logistic regression models, we only compare our tests with the $GRP_n$ test proposed by \cite{Jankova2020}. The simulation results are presented in Tables 9 and 10. We observe that the empirical sizes of our tests $\hat{T}_{\rm{min}}^{C}$ and $\hat{T}_{\rm{Fisher}}^{C}$ are close to the nominal significance level in most cases, whereas the test $GRP_n$ only occasionally maintains the significance level. In settings with high correlation and $p > n$, our tests $\hat{T}_{\rm{min}}^{C}$ and $\hat{T}_{\rm{Fisher}}^{C}$
exhibit inflated empirical sizes. In contrast, the empirical sizes of $GRP_n$ are unacceptably inflated even when the covariate dimension is small. In terms of empirical power, our proposed tests $\hat{T}_{\rm{min}}^{C}$ and $\hat{T}_{\rm{Fisher}}^{C}$ surpass the test $GRP_n$ in all scenarios where the latter successfully maintains the nominal size. Furthermore, similar to the results in Study 1, the empirical powers of our proposed tests increase as the correlation of the predictor vector $X$ grows.

\begin{table}[ht!]\caption{Empirical sizes and powers of our tests $\hat{T}_{\rm{min}}^C$ and $\hat{T}_{\rm{Fisher}}^C$ with different bandwidths for $H_{21}$ in Study 2.}\label{table-bandwidth-H21}
\centering
{\small\scriptsize\hspace{6cm}
\renewcommand{\arraystretch}{0.6}\tabcolsep 0.4cm
\begin{tabular}{*{20}{c}}
\hline
&\multicolumn{1}{c}{$c_h$}&\multicolumn{1}{c}{n=600}&\multicolumn{1}{c}{n=600}&\multicolumn{1}{c}{n=600}&\multicolumn{1}{c}{n=600}
&\multicolumn{1}{c}{n=600}&\multicolumn{1}{c}{n=600}\\
&&\multicolumn{1}{c}{p=50}&\multicolumn{1}{c}{p=100}&\multicolumn{1}{c}{p=300}&\multicolumn{1}{c}{p=600}
&\multicolumn{1}{c}{p=900}&\multicolumn{1}{c}{p=1200}\\
\hline
$\hat{T}_{\rm{min}}^C, \ \Sigma = I_p, \ a = 0$
& 0.75 &0.042 &0.030 &0.053 &0.059 &0.072 &0.061\\
& 1.00 &0.026 &0.032 &0.061 &0.049 &0.064 &0.055\\
& 1.25 &0.024 &0.021 &0.051 &0.059 &0.068 &0.064\\
\\
$\hat{T}^C_{\rm{Fisher}}, \ \Sigma = I_p, \ a = 0$
& 0.75 &0.036 &0.031 &0.031 &0.046 &0.045 &0.044\\
& 1.00 &0.031 &0.024 &0.042 &0.028 &0.032 &0.036\\
& 1.25 &0.026 &0.018 &0.025 &0.032 &0.048 &0.039\\
\\
$\hat{T}_{\rm{min}}^C, \ \Sigma = I_p, \ a = 1$
& 0.75 &0.610 &0.465 &0.240 &0.116 &0.116 &0.080\\
& 1.00 &0.634 &0.492 &0.222 &0.120 &0.087 &0.102\\
& 1.25 &0.634 &0.492 &0.222 &0.120 &0.087 &0.087\\
\\
$\hat{T}^C_{\rm{Fisher}}, \ \Sigma = I_p, \ a = 1$
& 0.75 &0.588 &0.482 &0.166 &0.084 &0.070 &0.064\\
& 1.00 &0.665 &0.509 &0.214 &0.085 &0.081 &0.075\\
& 1.25 &0.707 &0.542 &0.195 &0.096 &0.059 &0.064\\
\hline
$\hat{T}_{\rm{min}}^C, \ \Sigma = (0.4^{|i-j|})_{p\times p}, \ a = 0$
& 0.75 &0.024 &0.027 &0.060 &0.074 &0.091 &0.095\\
& 1.00 &0.025 &0.029 &0.036 &0.069 &0.092 &0.094\\
& 1.25 &0.022 &0.017 &0.027 &0.055 &0.074 &0.121\\
\\
$\hat{T}^C_{\rm{Fisher}}, \ \Sigma = (0.4^{|i-j|})_{p\times p}, \ a = 0$
& 0.75 &0.027 &0.028 &0.039 &0.052 &0.067 &0.067\\
& 1.00 &0.025 &0.029 &0.029 &0.050 &0.067 &0.073\\
& 1.25 &0.020 &0.015 &0.016 &0.034 &0.054 &0.092\\
\\
$\hat{T}_{\rm{min}}^C, \ \Sigma = (0.4^{|i-j|})_{p\times p}, \ a = 1$
& 0.75 &0.977 &0.929 &0.795 &0.641 &0.628 &0.512\\
& 1.00 &0.986 &0.962 &0.837 &0.696 &0.599 &0.565\\
& 1.25 &0.988 &0.958 &0.854 &0.755 &0.608 &0.577\\
\\
$\hat{T}^C_{\rm{Fisher}}, \ \Sigma = (0.4^{|i-j|})_{p\times p}, \ a = 1$
& 0.75 &0.986 &0.959 &0.782 &0.600 &0.572 &0.459\\
& 1.00 &0.990 &0.972 &0.827 &0.670 &0.548 &0.520\\
& 1.25 &0.995 &0.968 &0.839 &0.719 &0.563 &0.545\\
\hline
$\hat{T}_{\rm{min}}^C, \ \Sigma = (0.8^{|i-j|})_{p\times p}, \ a = 0$
& 0.75 &0.027 &0.037 &0.039 &0.069 &0.076 &0.110\\
& 1.00 &0.027 &0.030 &0.031 &0.050 &0.103 &0.118\\
& 1.25 &0.024 &0.025 &0.045 &0.054 &0.098 &0.133\\
\\
$\hat{T}^C_{\rm{Fisher}}, \ \Sigma = (0.8^{|i-j|})_{p\times p}, \ a = 0$
& 0.75 &0.034 &0.036 &0.037 &0.063 &0.078 &0.104\\
& 1.00 &0.026 &0.030 &0.029 &0.041 &0.092 &0.116\\
& 1.25 &0.029 &0.024 &0.035 &0.052 &0.104 &0.119\\
\\
$\hat{T}_{\rm{min}}^C, \ \Sigma = (0.8^{|i-j|})_{p\times p}, \ a = 1$
& 0.75 &0.999 &0.993 &0.933 &0.880 &0.868 &0.839\\
& 1.00 &1.000 &0.995 &0.954 &0.918 &0.884 &0.875\\
& 1.25 &1.000 &0.999 &0.956 &0.935 &0.893 &0.888\\
\\
$\hat{T}^C_{\rm{Fisher}}, \ \Sigma = (0.8^{|i-j|})_{p\times p}, \ a = 1$
& 0.75 &0.999 &0.997 &0.938 &0.873 &0.854 &0.835\\
& 1.00 &1.000 &0.996 &0.958 &0.919 &0.871 &0.857\\
& 1.25 &1.000 &1.000 &0.953 &0.938 &0.890 &0.875\\
\hline
\end{tabular}}
\end{table}

\begin{table}[ht!]\caption{Empirical sizes and powers of our tests $\hat{T}_{\rm{min}}^C$ and $\hat{T}^C_{\rm{Fisher}}$ with different bandwidths for $H_{22}$ in Study 2.}\label{table-bandwidth-H22}
\centering
{\small\scriptsize\hspace{8cm}
\renewcommand{\arraystretch}{0.6}\tabcolsep 0.4cm
\begin{tabular}{*{20}{c}}
\hline
&\multicolumn{1}{c}{$c_h$}&\multicolumn{1}{c}{n=600}&\multicolumn{1}{c}{n=600}&\multicolumn{1}{c}{n=600}
&\multicolumn{1}{c}{n=600}&\multicolumn{1}{c}{n=600}&\multicolumn{1}{c}{n=600}\\
&&\multicolumn{1}{c}{p=50}&\multicolumn{1}{c}{p=100}&\multicolumn{1}{c}{p=300}&\multicolumn{1}{c}{p=600}
&\multicolumn{1}{c}{p=900}&\multicolumn{1}{c}{p=1200}\\
\hline
$\hat{T}_{\rm{min}}^C, \ \Sigma = I_p, \ a = 0$
& 0.75 &0.041 &0.027 &0.049 &0.058 &0.056 &0.074\\
& 1.00 &0.026 &0.035 &0.043 &0.055 &0.067 &0.077\\
& 1.25 &0.015 &0.021 &0.045 &0.056 &0.060 &0.066\\
\\
$\hat{T}^C_{\rm{Fisher}}, \ \Sigma = I_p, \ a = 0$
& 0.75 &0.033 &0.030 &0.031 &0.043 &0.045 &0.047\\
& 1.00 &0.024 &0.027 &0.031 &0.029 &0.055 &0.054\\
& 1.25 &0.019 &0.018 &0.031 &0.028 &0.042 &0.042\\
\\
$\hat{T}_{\rm{min}}^C, \ \Sigma = I_p, \ a = 1$
& 0.75 &0.194 &0.141 &0.099 &0.064 &0.071 &0.065\\
& 1.00 &0.239 &0.161 &0.080 &0.080 &0.061 &0.060\\
& 1.25 &0.233 &0.167 &0.092 &0.061 &0.066 &0.074\\
\\
$\hat{T}^C_{\rm{Fisher}}, \ \Sigma = I_p, \ a = 1$
& 0.75 &0.218 &0.172 &0.065 &0.034 &0.056 &0.042\\
& 1.00 &0.267 &0.161 &0.063 &0.063 &0.041 &0.037\\
& 1.25 &0.271 &0.195 &0.071 &0.045 &0.038 &0.047\\
\hline
$\hat{T}_{\rm{min}}^C, \ \Sigma = (0.4^{|i-j|})_{p\times p}, \ a = 0$
& 0.75 &0.034 &0.031 &0.040 &0.073 &0.089 &0.111\\
& 1.00 &0.037 &0.030 &0.057 &0.068 &0.098 &0.088\\
& 1.25 &0.019 &0.033 &0.041 &0.065 &0.083 &0.119\\
\\
$\hat{T}^C_{\rm{Fisher}}, \ \Sigma = (0.4^{|i-j|})_{p\times p}, \ a = 0$
& 0.75 &0.038 &0.028 &0.027 &0.041 &0.072 &0.074\\
& 1.00 &0.032 &0.027 &0.040 &0.033 &0.063 &0.077\\
& 1.25 &0.019 &0.028 &0.023 &0.045 &0.055 &0.094\\
\\
$\hat{T}_{\rm{min}}^C, \ \Sigma = (0.4^{|i-j|})_{p\times p}, \ a = 1$
& 0.75 &0.835 &0.742 &0.599 &0.483 &0.433 &0.359\\
& 1.00 &0.882 &0.807 &0.672 &0.525 &0.452 &0.388\\
& 1.25 &0.902 &0.827 &0.689 &0.572 &0.477 &0.426\\
\\
$\hat{T}^C_{\rm{Fisher}}, \ \Sigma = (0.4^{|i-j|})_{p\times p}, \ a = 1$
& 0.75 &0.868 &0.782 &0.581 &0.447 &0.380 &0.320\\
& 1.00 &0.917 &0.860 &0.652 &0.484 &0.411 &0.329\\
& 1.25 &0.926 &0.868 &0.686 &0.519 &0.426 &0.375\\
\hline
\\
$\hat{T}_{\rm{min}}^C, \ \Sigma = (0.8^{|i-j|})_{p\times p}, \ a = 0$
& 0.75 &0.038 &0.027 &0.044 &0.070 &0.097 &0.130\\
& 1.00 &0.025 &0.027 &0.028 &0.064 &0.107 &0.138\\
& 1.25 &0.020 &0.026 &0.032 &0.059 &0.108 &0.111\\
\\
$\hat{T}^C_{\rm{Fisher}}, \ \Sigma = (0.8^{|i-j|})_{p\times p}, \ a = 0$
& 0.75 &0.033 &0.026 &0.035 &0.068 &0.084 &0.122\\
& 1.00 &0.028 &0.021 &0.022 &0.060 &0.094 &0.124\\
& 1.25 &0.021 &0.034 &0.039 &0.050 &0.089 &0.105\\
\\
$\hat{T}_{\rm{min}}^C, \ \Sigma = (0.8^{|i-j|})_{p\times p}, \ a = 1$
& 0.75 &1.000 &0.996 &0.951 &0.912 &0.898 &0.876\\
& 1.00 &0.997 &0.994 &0.964 &0.944 &0.915 &0.893\\
& 1.25 &1.000 &0.996 &0.979 &0.950 &0.933 &0.910\\
\\
$\hat{T}^C_{\rm{Fisher}}, \ \Sigma = (0.8^{|i-j|})_{p\times p}, \ a = 1$
& 0.75 &1.000 &0.996 &0.952 &0.914 &0.889 &0.859\\
& 1.00 &0.999 &0.995 &0.970 &0.947 &0.912 &0.871\\
& 1.25 &1.000 &1.000 &0.985 &0.949 &0.927 &0.903\\
\hline
\end{tabular}}
\end{table}

\begin{table}[ht!]\caption{Empirical sizes and powers of the tests $GRP_n$, $\hat{T}_{\rm{min}}^C$, and $\hat{T}_{\rm{Fisher}}^C$ with bandwidth $h = n^{-2/9}$ for $H_{21}$ in Study 2.}\label{table-H21}
\centering
{\small\scriptsize\hspace{8cm}
\renewcommand{\arraystretch}{0.6}\tabcolsep 0.4cm
\begin{tabular}{*{20}{c}}
\hline
&\multicolumn{1}{c}{a}&\multicolumn{1}{c}{n=600}&\multicolumn{1}{c}{n=600}&\multicolumn{1}{c}{n=600}&\multicolumn{1}{c}{n=600}
&\multicolumn{1}{c}{n=600}&\multicolumn{1}{c}{n=600}\\
&&\multicolumn{1}{c}{p=50}&\multicolumn{1}{c}{p=100}&\multicolumn{1}{c}{p=300}&\multicolumn{1}{c}{p=600}
&\multicolumn{1}{c}{p=900}&\multicolumn{1}{c}{p=1200}  \\
\hline
$\hat{T}_{\rm{min}}^C, \ \Sigma = I_p$
& 0.0 &0.022 &0.024 &0.054 &0.061 &0.068 &0.065\\
& 1.0 &0.589 &0.459 &0.229 &0.119 &0.086 &0.081\\
\\
$\hat{T}^C_{\rm{Fisher}}, \ \Sigma = I_p$
& 0.0 &0.017 &0.021 &0.027 &0.037 &0.047 &0.050\\
& 1.0 &0.648 &0.512 &0.214 &0.097 &0.067 &0.052\\
\\
$GRP_n, \ \Sigma = I_p$
& 0.0 &0.065 &0.036 &0.024 &0.025 &0.019 &0.024\\
& 1.0 &0.099 &0.061 &0.044 &0.027 &0.019 &0.017\\
\hline
$\hat{T}_{\rm{min}}^C, \ \Sigma = (0.4^{|i-j|})_{p\times p}$
& 0.0 &0.031 &0.028 &0.051 &0.061 &0.092 &0.090\\
& 1.0 &0.991 &0.943 &0.819 &0.722 &0.663 &0.537\\
\\
$\hat{T}^C_{\rm{Fisher}}, \ \Sigma = (0.4^{|i-j|})_{p\times p}$
& 0.0 &0.033 &0.023 &0.034 &0.043 &0.064 &0.070\\
& 1.0 &0.997 &0.958 &0.812 &0.667 &0.609 &0.489\\
\\
$GRP_n, \ \Sigma = (0.4^{|i-j|})_{p\times p}$
& 0.0 &0.198 &0.179 &0.130 &0.092 &0.087 &0.069\\
& 1.0 &0.770 &0.490 &0.172 &0.102 &0.072 &0.062\\
\hline
$\hat{T}_{\rm{min}}^C, \ \Sigma = (0.8^{|i-j|})_{p\times p}$
& 0.0 &0.023 &0.026 &0.037 &0.055 &0.086 &0.118\\
& 1.0 &0.999 &0.994 &0.942 &0.879 &0.897 &0.874\\
\\
$\hat{T}^C_{\rm{Fisher}}, \ \Sigma = (0.8^{|i-j|})_{p\times p}$
& 0.0 &0.025 &0.031 &0.032 &0.048 &0.079 &0.113\\
& 1.0 &0.999 &0.996 &0.944 &0.874 &0.879 &0.860\\
\\
$GRP_n, \ \Sigma = (0.8^{|i-j|})_{p\times p}$
& 0.0 &0.422 &0.373 &0.362 &0.327 &0.306 &0.314\\
& 1.0 &1.000 &1.000 &0.955 &0.980 &0.935 &0.878\\
\hline
\end{tabular}}
\end{table}

\begin{table}[ht!]\caption{Empirical sizes and powers of the tests  $GRP_n$, $\hat{T}_{\rm{min}}^C$, and $\hat{T}_{\rm{Fisher}}^C$ with bandwidth $h = n^{-2/9}$ for $H_{22}$ in Study 2.}\label{table-H22}
\centering
{\small\scriptsize\hspace{8cm}
\renewcommand{\arraystretch}{0.6}\tabcolsep 0.4cm
\begin{tabular}{*{20}{c}}
\hline
&\multicolumn{1}{c}{a}&\multicolumn{1}{c}{n=600}&\multicolumn{1}{c}{n=600}&\multicolumn{1}{c}{n=600}&\multicolumn{1}{c}{n=600}
&\multicolumn{1}{c}{n=600}&\multicolumn{1}{c}{n=600}\\
&&\multicolumn{1}{c}{p=50}&\multicolumn{1}{c}{p=100}&\multicolumn{1}{c}{p=300}&\multicolumn{1}{c}{p=600}
&\multicolumn{1}{c}{p=900}&\multicolumn{1}{c}{p=1200}  \\
\hline
$\hat{T}_{\rm{min}}^C, \ \Sigma = I_p$
& 0.0 &0.031 &0.031 &0.058 &0.053 &0.063 &0.061\\
& 1.0 &0.199 &0.174 &0.095 &0.074 &0.059 &0.078\\
\\
$\hat{T}^C_{\rm{Fisher}}, \ \Sigma = I_p$
& 0.0 &0.031 &0.026 &0.032 &0.032 &0.042 &0.041\\
& 1.0 &0.220 &0.194 &0.080 &0.041 &0.041 &0.052\\
\\
$GRP_n, \ \Sigma = I_p$
& 0.0 &0.069 &0.054 &0.040 &0.021 &0.022 &0.014\\
& 1.0 &0.224 &0.107 &0.040 &0.026 &0.023 &0.017\\
\hline
$\hat{T}_{\rm{min}}^C, \ \Sigma = (0.4^{|i-j|})_{p\times p}$
& 0.0 &0.034 &0.027 &0.042 &0.078 &0.104 &0.096\\
& 1.0 &0.869 &0.792 &0.655 &0.528 &0.481 &0.377\\
\\
$\hat{T}^C_{\rm{Fisher}}, \ \Sigma = (0.4^{|i-j|})_{p\times p}$
& 0.0 &0.029 &0.031 &0.030 &0.054 &0.079 &0.072\\
& 1.0 &0.918 &0.840 &0.658 &0.472 &0.417 &0.322\\
\\
$GRP_n, \ \Sigma = (0.4^{|i-j|})_{p\times p}$
& 0.0 &0.203 &0.208 &0.123 &0.105 &0.077 &0.081\\
& 1.0 &0.947 &0.826 &0.376 &0.227 &0.141 &0.105\\
\hline
$\hat{T}_{\rm{min}}^C, \ \Sigma = (0.8^{|i-j|})_{p\times p}$
& 0.0 &0.027 &0.029 &0.036 &0.051 &0.087 &0.117\\
& 1.0 &1.000 &0.996 &0.968 &0.945 &0.925 &0.880\\
\\
$\hat{T}^C_{\rm{Fisher}}, \ \Sigma = (0.8^{|i-j|})_{p\times p}$
& 0.0 &0.029 &0.029 &0.029 &0.048 &0.077 &0.100\\
& 1.0 &1.000 &0.995 &0.975 &0.942 &0.917 &0.861\\
\\
$GRP_n, \ \Sigma = (0.8^{|i-j|})_{p\times p}$
& 0.0 &0.402 &0.389 &0.351 &0.343 &0.297 &0.331\\
& 1.0 &1.000 &1.000 &0.960 &0.990 &0.988 &0.963\\
\hline
\end{tabular}}
\end{table}

\subsection{Real data examples}
\begin{figure}[t!]
\centering
\includegraphics[width=0.4 \textwidth]{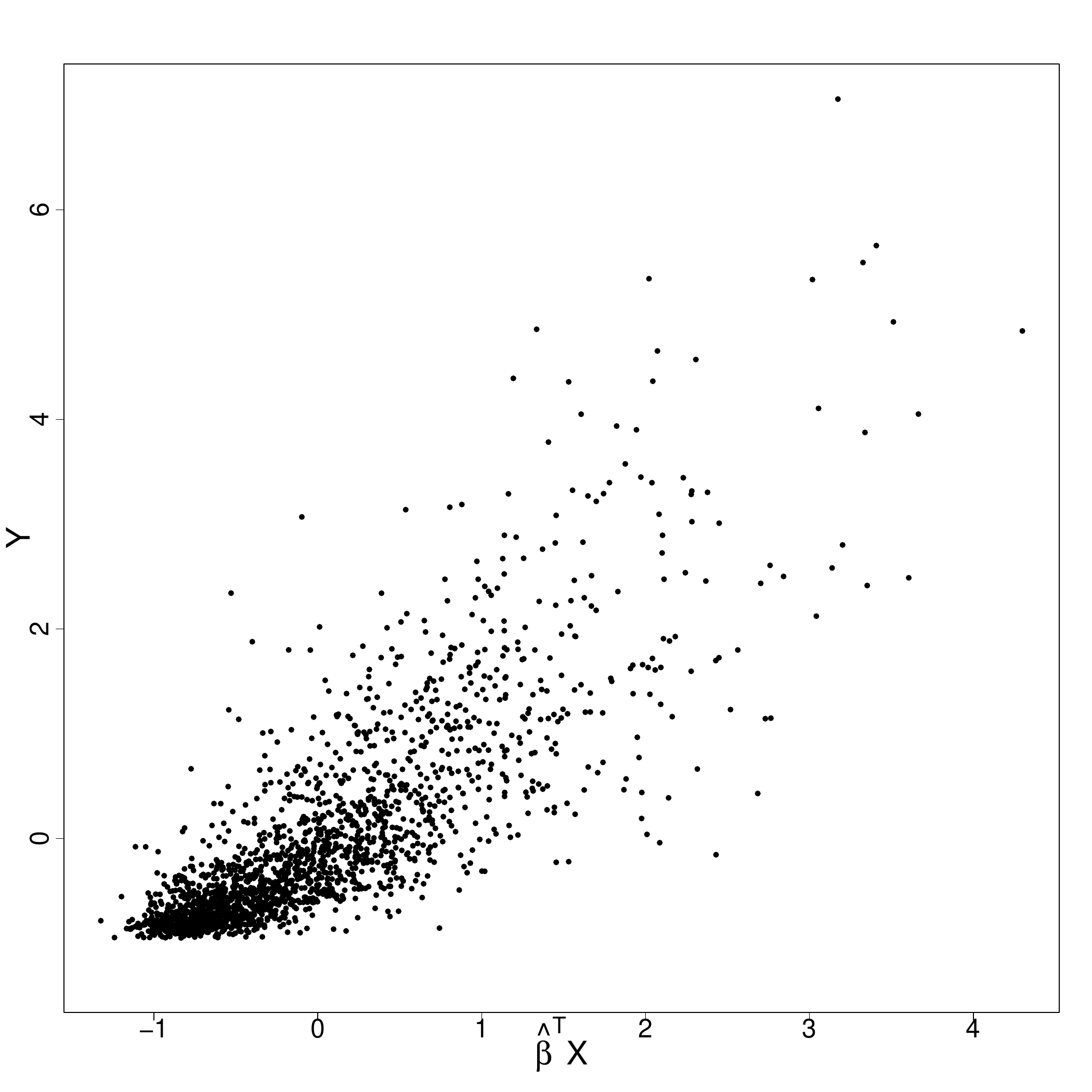}
\caption{Scatterplot of $Y$ versus the projected covariate $\hat{\beta}^{\top}X$.}
\label{fig_realdata_linear}
\end{figure}

\begin{figure}[t!]
\centering
\includegraphics[width=0.4 \textwidth]{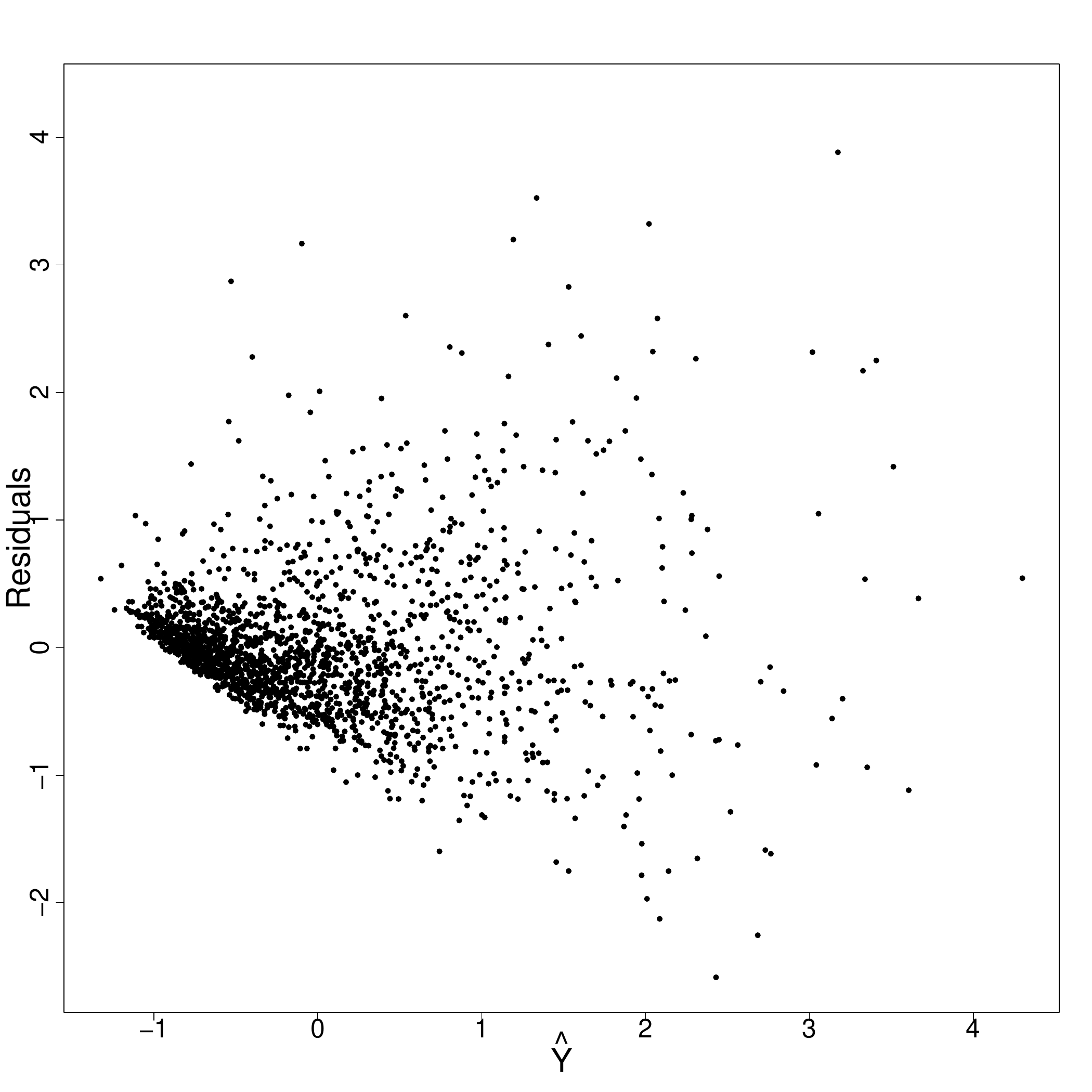}
\caption{Scatterplot of Residuals from the polynomial regression model versus the fitted values $\hat{Y}$.}
\label{fig_realdata_nonlinear}
\end{figure}

In this subsection, we evaluate the proposed tests using two real data analyses. The first is a regression task on the Communities and Crime Unnormalized dataset from 1995 in the US. This dataset is available at
\url{https://archive.ics.uci.edu/dataset/211/communities+and+crime+unnormalized}. It contains 2215 observations collected by combining socio-economic data from the '90 Census, law enforcement data from the 1990 Law Enforcement Management and Admin Stats survey, and crime data from the 1995 FBI UCR. Each observation has 125 features and 18 potential target variables. To apply our tests, we first clean the data by removing observations with missing values, leaving 1901 observations and 102 predictive features. The response variable $Y$ we use is total crime, and we let $X=(X_1, X_2, \dots, X_{102})^{\top}$ represent the predictor vector. All variables are standardized separately for ease of interpretation. We then apply our tests $\hat{T}_{\rm{min}}^C$ and $\hat{T}_{\rm{Fisher}}^C$ to check whether a linear regression model is adequate for fitting this data. The choices of the bandwidth $h$ and projections are the same as in the simulation studies. The $p$-values of $\hat{T}_{\rm{min}}^C$ and $\hat{T}_{\rm{Fisher}}^C$ are approximately $ 0.00002$ and $0.0003$, respectively, which indicates a rejection of the linearity of the underlying regression model. We present a scatterplot of $Y$ against $\hat{\beta}^{\top}X$ in Figure \ref{fig_realdata_linear}, where $\hat{\beta}$ is the estimator from the linear model. This plot also suggests that a linear relationship between $Y$ and $X$ may not be plausible. It appears that there exists a quadratic relationship between $Y$ and $\hat{\beta}^{\top}X$. Therefore, we consider a polynomial regression model by incorporating linear, squared, and interaction terms for the predictors. When the proposed tests $\hat{T}_{\rm{min}}^C$ and $\hat{T}_{\rm{Fisher}}^C$ are applied to this polynomial regression model, the corresponding $p$-values are $0.389$ and $0.338$, respectively. This suggests that the polynomial regression model with quadratic and interaction terms may be plausible. To further visualize this fit, Figure \ref{fig_realdata_nonlinear} presents a scatterplot of the residuals from this polynomial regression model against the fitted values $\hat{Y}$. We can see that there is no clear trend between the residuals and the fitted values. Therefore, this polynomial regression model is plausible.

The second analysis is a classification task that aims at predicting whether a molecule is musk or non-musk. The dataset is available at \url{https://archive.ics.uci.edu/dataset/74/musk+version+1}. It contains 476 observations and 168 features. We first remove non-predictive variables (``molecule name'' and ``conformation name''), leaving 476 observations and 166 features. The label value $Y$ represents the category of the molecule (0: non-musk, 1: musk). We then check whether a sparse linear logistic regression model is adequate for a classification task and apply the proposed tests $\hat{T}_{\rm{min}}^C$ and $\hat{T}_{\rm{Fisher}}^C$ to check whether $E(Y|X)=\exp{(\beta^T X)}/(1+\exp{(\beta^T X)})$ is plausible. The choices for the bandwidth $h$ and projections are also the same as in the simulation studies. The $p$-values of $\hat{T}_{\rm{min}}^C$ and $\hat{T}_{\rm{Fisher}}^C$ are approximately $0.608$ and $0.666$, respectively, which indicates that a linear logistic regression model may be adequate to fit this data. We further calculate the predictive accuracy of the linear logistic regression model. We conducted $50$ runs to reduce the bias resulting from the randomness in selecting samples for the training and testing sets. In each run, the dataset was randomly shuffled and split into a training set ($75\%$) and a testing set ($25\%$). The models are estimated using the training set, then the predictive accuracy and the Area Under the Curve (AUC) are computed using the testing set. We then obtain an average predictive accuracy of $80.4\%$ and an AUC of $87.5\%$ for the linear logistic regression model, which further confirms that the linear logistic regression model may be suitable for fitting this dataset.

\section{Conclusion}
In this paper, we develop a new methodology for testing the goodness-of-fit of sparse regression models based on projections, when the covariate dimension $p$ may substantially exceed the sample size $n$. Most existing goodness-of-fit tests for regressions cannot typically be extended to ultra-high dimensional settings due to challenges arising from the curse of dimensionality or dependencies on the asymptotic linearity and normality of parameter estimators. In contrast, our projection-based tests do not rely on the asymptotic expansion or normality of high dimensional parameter estimators. We investigated the asymptotic properties of the projection-based test statistics under the null and alternative hypotheses, where the growing rate of the dimension $p$ is of exponential order in relation to the sample size. Under the alternative hypothesis, the projection-based test is consistent with asymptotic power 1 for almost all projections on the unit sphere and can detect local alternatives departing from the null at the rate of $O(n^{-1/2} h^{-1/4})$. Notably, this detection rate is independent of the covariate dimension $p$ even in ultra-high dimensional settings. An interesting theoretical result is that for pairwise linearly independent projections, the resulting projection-based test statistics are asymptotically independent. This result can also be applied to other statistical problems via projections, such as high-dimensional significance testing. We then propose two combined tests based on combination methods and data splitting. Since the test statistics only involve one-dimensional nonparametric smoothing, our tests significantly mitigate the curse of dimensionality, even when the covariate dimension substantially exceeds the sample size. The simulation results in Section 5 align with these theoretical results in finite samples. However, employing data splitting introduces variability in the values of the test statistics. It would be interesting to develop goodness-of-fit tests in ultra-high dimensional settings where the data splitting strategy is completely avoided.

\section*{Acknowledgement}
Heng Peng's research was supported by the University Grants Council of Hong Kong. Lixing Zhu's research was supported by the  grants  (NSFC12131006, NSFC12471276) from the National Natural Scientific Foundation of China  and the grant (CI2023C063YLL) from the Scientific and Technological Innovation Project of China Academy of Chinese Medical Science.

\clearpage
\bibliographystyle{apalike}
\bibliography{sparse_glm}

\end{document}